\documentclass[10pt,journal]{IEEEtran}
\usepackage{lineno,hyperref}
\usepackage{amssymb}
\usepackage{mathtools, cuted}
\usepackage{color, comment, verbatim}
\usepackage{latexsym}
\usepackage{enumerate}
\usepackage{caption}
\usepackage{subcaption}
\usepackage{algorithm, algpseudocode}
\usepackage{graphics}
\usepackage{graphicx}
\usepackage[dvipsnames]{xcolor}
\newtheorem{lemma}{Lemma}
\newtheorem{theorem}{Theorem}
\ifCLASSINFOpdf
\else
\fi

\begin{document}
%
\title{Robust Information Criterion for Model Selection in Sparse High-Dimensional Linear Regression Models}
%
%

\author{Prakash~B. Gohain,~\IEEEmembership{Student Member,~IEEE,}
        Magnus~Jansson,~\IEEEmembership{Senior Member,~IEEE.}
\thanks{This research was supported in part by the European Research Council (ERC) under the European Union’s Horizon 2020 research and innovation programme, grant agreement No.  742648. 

The authors are with the Division of Information Science and Engineering, KTH Royal Institute of Technology, Stockholm SE-10044, Sweden
e-mail: pbg@kth.se, janssonm@kth.se.}
}
%
%

\markboth{IEEE transactions on signal processing, 2022}%
{Shell \MakeLowercase{\textit{et al.}}: Bare Demo of IEEEtran.cls for IEEE Journals}
%



\maketitle

\begin{abstract}
Model selection in linear regression models is a major challenge when dealing with high-dimensional data where the number of available measurements (sample size) is much smaller than the dimension of the parameter space. Traditional methods for model selection such as Akaike information criterion, Bayesian information criterion (BIC) and minimum description length are heavily prone to overfitting in the high-dimensional setting. In this regard, extended BIC (EBIC), which is an extended version of the original BIC and extended Fisher information criterion (EFIC), which is a combination of EBIC and Fisher information criterion, are consistent estimators of the true model as the number of measurements grows very large. However, EBIC is not consistent in high signal-to-noise-ratio (SNR) scenarios where the sample size is fixed and EFIC is not invariant to data scaling resulting in unstable behaviour. In this paper, we propose a new form of the EBIC criterion called EBIC-Robust, which is invariant to data scaling and consistent in both large sample size and high-SNR scenarios. Analytical proofs are presented to guarantee its consistency. Simulation results indicate that the performance of EBIC-Robust is quite superior to that of both EBIC and EFIC.

\end{abstract}


%
\IEEEpeerreviewmaketitle

\section{Introduction}
Selecting the true or best set of covariates from a large pool of potential covariates is a fundamental requirement in many applications of science, engineering and biology. In this paper, our primary focus is on model selection in high-dimensional linear regression models associated with the maximum likelihood (ML) method of parameter estimation where the number of measurements, $N$, is quite small compared to the model space or parameter dimension, $p$, i.e., $N < p$. High-dimensional datasets are a common phenomena in many fields of scientific studies, and as such model selection is a central element of data analysis and statistical inference \cite{MOS_overview_2018}. 

Consider the linear model
\begin{equation}
\mathbf{y} = \mathbf{A}\mathbf{x} + \mathbf{e}, \label{eq:LR_model_1}
\end{equation}
where $\mathbf{y} \in \mathbb{R}^{N}$ is the measurement vector and $\mathbf{A}\in \mathbb{R}^{N\times p}$ is the known design matrix. We are considering a high-dimensional setting, hence $p> N$. Also, $p$ can be linked to $N$ as $p=N^d$, where $d>0$ is a real value. $\mathbf{e} \in \mathbb{R}^{N}$ is the associated noise vector whose elements are assumed to be i.i.d. following a Gaussian distribution, i.e., $\mathbf{e} \sim \mathcal{N}(\mathbf{0},\sigma^2\mathbf{I}_N)$ where $\sigma^2$ is the unknown true noise power. $\mathbf{x} \in \mathbb{R}^{p}$ is the unknown parameter vector. Here, $\mathbf{x}$ is assumed to be sparse, which implies that very few of the elements of $\mathbf{x}$ are non-zero. We denote $\mathcal{S}$ as the true support of $\mathbf{x}$, i.e., $\mathcal{S} = \{i : x_i \neq 0\}$ having cardinality $card(\mathcal{S}) = k_0 \ll N$ and $\mathbf{A}_\mathcal{S}$ as the set of columns of $\mathbf{A}$ corresponding to the support $\mathcal{S}$. The goal of model selection is estimating $\mathcal{S}$ given $\mathbf{y}$ and $\mathbf{A}$.

A popular approach for model selection is using information theoretic criteria \cite{MOS_review_Stoica2004,MS_Rao_2001,MS_review_Anderson_2004,MS_review_chakrbarti_2011}. A typical information criterion based model selection rule picks the best model that minimizes some statistical metric as shown below
\begin{equation}
    \hat{\mathcal{S}} = \underset{\mathcal{I}\in \mathcal{J} }{\arg \min}  \{ f(\mathcal{M}_\mathcal{I}) + \mathcal{P}(\mathcal{I})  \},
\end{equation}
where $\hat{\mathcal{S}}$ is the model estimate, $\mathcal{J}$ is the set of candidate models under consideration and $\mathcal{M}_\mathcal{I}$ denotes the model with support $\mathcal{I}$. The statistical metric consists of two parts: (1) $f(\mathcal{M}_\mathcal{I})$ representing the goodness of fit of model $\mathcal{M}_\mathcal{I}$ and (2) $\mathcal{P}(\mathcal{I})$ is the penalty term that compensates for overparameterization.
The literature on model selection is quite extensive. Some of the popular classical model selection rules include Akaike information criterion \cite{AIC_Akaike1974}, Bayesian information criterion (BIC)\cite{BIC_Schwarz1978}, minimum description length (MDL)\cite{MDL1978}, gMDL\cite{gMDL_Hansen2001}, nMDL\cite{NML_Rissanen2000}, and penalizing adaptively the likelihood (PAL) \cite{PAL_Stoica2013}.
However, these classical methods in their current form fail to handle the large dimension cases and tend to overfit the final model \cite{EBIC2008,EFIC2018}.

Among the classical methods of model selection, BIC has been quite successful due to its simplicity and consistent performance in many fields. BIC is asymptotically consistent in selecting the true model as $N$ grows very large given that $p$ and the true noise variance $\sigma^2$ is fixed. However, its performance in high-dimensional settings when $p > N$ is not satisfactory and it has a tendency to select more co-variates than required, thus overfitting the model \cite{EBIC2008}. To handle the large-$p$ small-$N$ scenario, the authors in \cite{EBIC2008} proposed a novel extension to the original BIC called extended BIC (EBIC), that takes into account both the number of unknown parameters and the complexity of the model space. EBIC adds dynamic prior model probabilities to each of the models under consideration that is inversely proportional to the model set dimension. This eliminates the earlier assumption of assigning uniform prior to all models irrespective of their sizes, which goes against the principle of parsimony. Under a suitable asymptotic identifiability condition, EBIC is consistent such that it selects the true model as $N$ tends to infinity \cite{EBIC2008}. However, the consistent behaviour of EBIC fails when $N$ is small and fixed and $\sigma^2$  tends to zero \cite{EFIC2018}. This new consistency requirement was first introduced in \cite{EEF_Kay2005}, where the authors highlighted that the original BIC is also inconsistent for fixed $N$ and decreasing noise variance scenarios where $N>p$.

To overcome the drawbacks of EBIC, the authors in \cite{EFIC2018} proposed a novel criterion called extended Fisher information criterion (EFIC) that is inspired by EBIC and the model selection criteria with Fisher information \cite{FIC}.
The authors analyzed the performance of
EFIC in the high-dimensional setting for two key cases: (1) when $\sigma^2$ is fixed and $N$ tends to infinity; (2) when $N$ is fixed and $\sigma^2$ tends to zero. In each case, it was shown that EFIC selects the true model with a probability approaching one. 
However, as indicated in our simulations, EFIC is not invariant to data scaling and it tends to suffer from overfitting issues (and sometimes underfitting) in practical sizes of $N$ when the data is scaled. This scaling problem is a result of the data dependent penalty design that may blow the penalty to extremely small or large values depending on how the data is scaled.

Apart from the criteria mentioned above, there are other non-information theoretic methods available for model selection. One such popular method is cross-validation (CV) \cite{CV_picard1984}. However, the performance of CV is quite poor in sample scarce scenarios with large parameter dimensions and even though CV is unbiased, it can have high variance \cite{CV_p_greater_N}.
Recent additions to the list of model selection methods for high-dimensional data are residual ratio thresholding (RRT)\cite{RRT-2018}  and multi-beta-test (MBT) \cite{gohain2020_MBT}. Both are non-information theoretic methods based on hypothesis testing using a test statistic. They operate along with a greedy variable selection method such as orthogonal matching pursuit (OMP) \cite{OMP_Cai2011} and involve a tuning parameter $\in [0,1]$, that controls the false-discovery rate. However, there is no optimal way to set it and as such their behaviour may tend to overfit or underfit depending on the chosen tuning parameter value. Moreover, in their current form, they can only be used with algorithms that generate monotonic sequences of support estimates such as OMP, which restricts their usability. 

In this paper, we propose a modified criterion for model selection in high-dimensional linear regression models called EBIC-Robust or EBIC$_\text{R}$ in short, where the subscript R stands for robust. EBIC$_\text{R}$ is a scale-invariant and consistent criterion. To guarantee the consistency, we provide analytical proofs to show that under a suitable asymptotic identifiability condition, EBIC$_\text{R}$ selects the true model with a probability approaching one as $N\to \infty $ as well as when $\sigma^2 \to 0$. Some preliminary results have been published in \cite{Gohain2022EBICR}.

Throughout the paper, boldface letters denote matrices and vectors. The notation $(\cdot)^T$ stands for transpose.
$\mathbf{A}_\mathcal{I}$ denotes a sub-matrix of the full matrix $\mathbf{A}$ formed using the columns indexed by the support set $\mathcal{I}$.
$\boldsymbol{\Pi}_\mathcal{I} = \mathbf{A}_\mathcal{I}(\mathbf{A}_\mathcal{I}^T\mathbf{A}_\mathcal{I})^{-1}\mathbf{A}_\mathcal{I}^T$ denotes the orthogonal projection matrix on the span of $\mathbf{A}_\mathcal{I}$ and $\boldsymbol{\Pi}_\mathcal{I}^\perp = \mathbf{I}_N-\boldsymbol{\Pi}_\mathcal{I}$ denotes the orthogonal projection matrix on the null space of $\mathbf{A}_\mathcal{I}^T$ and $\mathbf{I}_N$ is an $N\times N$ identity matrix.
The notation $\big\lvert \mathbf{X} \big\rvert$ denotes the determinant of the matrix $\mathbf{X}$ and $\lVert \cdot \rVert_2$ denotes the Euclidean norm. $X\sim \mathcal{N}(0,1)$ denotes a normal distributed random variable with mean $0$ and variance $1$. 
$X \sim \chi^2_k$ is a central chi-squared distributed random variable with $k$ degrees of freedom, $X \sim \chi^2_k(\lambda)$ is a noncentral chi-squared distributed random variable with $k$ degrees of freedom and non-centrality parameter $\lambda$. 


\section{Background}
Given the linear model (\ref{eq:LR_model_1}), the entire process of model selection or in other words estimating the true support set $\mathcal{S}$ involves two major steps: (i) Predictor/subset selection, which includes finding a competent set of candidate models out of all the ($2^p-1$) possible models. In our work, we consider the set of competing models as the collection of all plausible combinatorial models up to a maximum cardinality $K$, under the assumption that $k_0\leq K \ll N$; (ii) estimating the true model among the candidate models using a suitable model selection criterion. For a candidate model with support $\mathcal{I}$ having cardinality $card(\mathcal{I}) = k$, the linear model in (\ref{eq:LR_model_1}) can be reformulated as follows
\begin{equation}
\mathcal{H}_\mathcal{I} : \mathbf{y} = \mathbf{A}_\mathcal{I}\mathbf{x}_\mathcal{I} + \mathbf{e}_\mathcal{I}, \label{eq:LR_model_2}
\end{equation}
where $\mathcal{H}_\mathcal{I}$  denotes the hypothesis that the data $\mathbf{y}$ is truly generated according to (\ref{eq:LR_model_2}), $\mathbf{A}_\mathcal{I}\in \mathbb{R}^{N\times k}$ is the sub-design matrix consisting of columns from the known design matrix $\mathbf{A}$ with support $\mathcal{I}$, $\mathbf{x}_\mathcal{I} \in \mathbb{R}^{k}$ is the corresponding unknown parameter vector and $\mathbf{e}_\mathcal{I} \in \mathbb{R}^{N}$ is the associated noise vector following $\mathbf{e}_\mathcal{I} \sim \mathcal{N}(\mathbf{0},\sigma^2_\mathcal{I}\mathbf{I}_N)$ where $\sigma^2_\mathcal{I}$ is the unknown noise variance corresponding to the hypothesis $\mathcal{H}_\mathcal{I}$.
\subsection{Bayesian Framework for Model Selection}
To motivate the proposed criterion we 
start by describing the Bayesian framework that leads to the maximum a-posteriori (MAP) estimator, which in turn forms the backbone for deriving BIC and its extended versions, viz., EBIC, EFIC, as well as the proposed criterion EBIC$_\text{R}$. Now, for the considered model in (\ref{eq:LR_model_2}), the probability density function (pdf) of the data vector $\mathbf{y}$ is given as
\begin{equation}
    p(\mathbf{y}\lvert \boldsymbol{\theta}_\mathcal{I},\mathcal{H}_\mathcal{I}) = \frac{ \exp\{- \lVert \mathbf{y} -\mathbf{A}_\mathcal{I}\mathbf{x}_\mathcal{I} \rVert^2_2/2\sigma^2_\mathcal{I}\} }{(2\pi \sigma^2_\mathcal{I})^{N/2}}, \label{eq:pdf_y}
\end{equation}
where $\boldsymbol{\theta}_\mathcal{I} = [\mathbf{x}^T_\mathcal{I}, \sigma^2_\mathcal{I}]^T$ comprises of all the parameters of the model. Under hypothesis $\mathcal{H}_\mathcal{I}$, the maximum likelihood estimates (MLEs) of $\hat{\boldsymbol{\theta}}_\mathcal{I} = [\hat{\mathbf{x}}^T_\mathcal{I}, \hat{\sigma}^2_\mathcal{I}]^T$
are obtained as \cite{S_Kay_estimation_book}
\begin{equation}
    \hat{\mathbf{x}}_\mathcal{I} = \left(\mathbf{A}^T_\mathcal{I}\mathbf{A}_\mathcal{I}\right)^{-1}\mathbf{A}^T_\mathcal{I} \mathbf{y} \quad \& \quad  \hat{\sigma}^2_\mathcal{I}=\frac{\mathbf{y}^T\mathbf{\Pi}^\perp_\mathcal{I}\mathbf{y}}{N}.\label{eq:MLEs_theta}
\end{equation}
Let $p(\boldsymbol{\theta}_\mathcal{I}|\mathcal{H}_\mathcal{I})$ denote the prior pdf of the parameter vector $\boldsymbol{\theta}_\mathcal{I}$ under $\mathcal{H}_\mathcal{I}$. Then we have the joint probability
\begin{equation}
    p(\mathbf{y},\boldsymbol{\theta}_\mathcal{I}|\mathcal{H}_\mathcal{I}) = p(\mathbf{y}|\boldsymbol{\theta}_\mathcal{I},\mathcal{H}_\mathcal{I})p(\boldsymbol{\theta}_\mathcal{I}|\mathcal{H}_\mathcal{I}) \label{eq:joint_density}
\end{equation}
and the marginal distribution of $\mathbf{y}$ is
\begin{equation}
    p(\mathbf{y}|\mathcal{H}_\mathcal{I}) = \int p(\mathbf{y}|\boldsymbol{\theta}_\mathcal{I},\mathcal{H}_\mathcal{I})p(\boldsymbol{\theta}_\mathcal{I}|\mathcal{H}_\mathcal{I})d\boldsymbol{\theta}_\mathcal{I}. \label{eq:marginal}
\end{equation}
The posterior probability $\Pr(\mathcal{H}_\mathcal{I}|\mathbf{y})$ is given by
\begin{equation}
    \Pr(\mathcal{H}_\mathcal{I}|\mathbf{y}) = \frac{p(\mathbf{y}|\mathcal{H}_\mathcal{I})\Pr\left(\mathcal{H}_\mathcal{I}\right)}{p(\mathbf{y})},\label{eq:posterior_prob}
\end{equation}
where $\Pr(\mathcal{H}_\mathcal{I})$ is the prior probability of the model with support $\mathcal{I}$. The MAP estimator picks the model with the largest posterior probability $\Pr(\mathcal{H}_\mathcal{I}|\mathbf{y})$.
However, note that the $p(\mathbf{y})$ is a normalizing factor and independent of $\mathcal{I}$. Hence, the MAP estimate of $\mathcal{S}$ is equivalently given by 
\begin{equation}
    \hat{\mathcal{S}}_\text{MAP} = \underset{\mathcal{I}}{\arg \max}\ \Big \{ \ln p(\mathbf{y}|\mathcal{H}_\mathcal{I}) + \ln \Pr\left(\mathcal{H}_\mathcal{I}\right) \Big \}.
\end{equation}
To compute the MAP estimate, we need to evaluate the integral in (\ref{eq:marginal}). Traditionally, under the assumption that $N$ and/or SNR are large, we can obtain an approximation of  $\ln p(\mathbf{y}|\mathcal{H}_\mathcal{I})$ using a second order Taylor series expansion, which gives (see \cite{Stoica_BIC_SNR,GOHAIN-BIC-R} for details)
\begin{equation}
    \begin{split}
    \ln p(\mathbf{y}|\mathcal{H}_\mathcal{I}) \approx  \ln p(\mathbf{y}|\hat{\boldsymbol{\theta}}_\mathcal{I},\mathcal{H}_\mathcal{I})+ \ln p(\hat{\boldsymbol{\theta}}_\mathcal{I}|\mathcal{H}_\mathcal{I})\\
    +\frac{k+1}{2}\ln(2\pi)-\frac{1}{2}\ln \big|\hat{\mathbf{F}}_\mathcal{I}\big|, \label{eq:approx_ln_pdf_of_y}
    \end{split}
\end{equation}
where $k={card}(\mathcal{I})$ and $\hat{\mathbf{F}}_\mathcal{I}$ is the sample Fisher information matrix under $\mathcal{H}_\mathcal{I}$ given as \cite{S_Kay_estimation_book}
\begin{equation}
    \hat{\mathbf{F}}_\mathcal{I} = -\frac{\partial^2\ln p(\mathbf{y}|{\boldsymbol{\theta}}_\mathcal{I},\mathcal{H}_\mathcal{I})}{\partial \boldsymbol{\theta}_\mathcal{I}\partial \boldsymbol{\theta}_\mathcal{I}^T}\bigg \lvert_{{\boldsymbol{\theta}}_\mathcal{I} = \hat{\boldsymbol{\theta}}_\mathcal{I}}.\label{eq:FIM_raw}
\end{equation}
Evaluating (\ref{eq:FIM_raw}) using (\ref{eq:pdf_y}) and (\ref{eq:MLEs_theta}) we get \cite{Stoica_BIC_SNR}
\begin{equation}
    \hat{\mathbf{F}}_\mathcal{I} = 
    \begin{bmatrix}
    \frac{1}{\hat{\sigma}^2_\mathcal{I}}\mathbf{A}_\mathcal{I}^T\mathbf{A}_\mathcal{I} & \mathbf{0} \\
    \mathbf{0} & \frac{N}{2\hat{\sigma}^4_\mathcal{I}}
    \end{bmatrix}.\label{eq:FIM_matrix}
\end{equation}
Now, for the considered linear model we have
\begin{equation}
    -2\ln p(\mathbf{y}|\boldsymbol{\hat{\theta}}_\mathcal{I},\mathcal{H}_\mathcal{I}) = N\ln \hat{\sigma}^2_\mathcal{I} +\text{const}.\label{eq:pdf_LR_MLE}
\end{equation}
Therefore, using (\ref{eq:pdf_LR_MLE}), we can rewrite (\ref{eq:approx_ln_pdf_of_y}) as
\begin{equation}
    \begin{split}
     -2\ln p(\mathbf{y}|\mathcal{H}_\mathcal{I}) \approx N\ln \hat{\sigma}^2_\mathcal{I} + \ln \big|\hat{\mathbf{F}}_\mathcal{I}\big | - 2\ln p(\hat{\boldsymbol{\theta}}_\mathcal{I}|\mathcal{H}_\mathcal{I})\\
     - k\ln 2\pi +\text{const.}
    \end{split}
\end{equation}
Furthermore, it is assumed that the prior term in (\ref{eq:approx_ln_pdf_of_y}), i.e., $\ln p(\hat{\boldsymbol{\theta}}_\mathcal{I}|\mathcal{H}_\mathcal{I})$ is flat and uninformative, and hence disregarded from the analysis. Thus, dropping the constants and the terms independent of the model dimension $k$, we can equivalently reformulate the MAP based model estimate as
\begin{equation}
     \hat{\mathcal{S}}_\text{MAP} = \underset{\mathcal{I}}{\arg \min}\Big\{N\ln \hat{\sigma}^2_\mathcal{I} + \ln \big| \hat{\mathbf{F}}_\mathcal{I}\big| -k\ln 2\pi -2\ln \Pr\left(\mathcal{H}_\mathcal{I}\right)  \Big \}.\label{eq:MAP_estimate_final}
\end{equation}

\subsection{BIC}
The BIC can be obtained from the MAP estimator in (\ref{eq:MAP_estimate_final}).
The term $-k\ln 2\pi$ is ignored as it weakly depends on the model dimension $k$ and hence is typically much smaller than the dominating terms. Moreover, the prior probability of each candidate model is assumed to be equiprobable. Hence, the $-2\ln \Pr(\mathcal{H}_\mathcal{I})$ term is dropped as well. Now, expanding the  $|\hat{\mathbf{F}}_\mathcal{I}|$ term of (\ref{eq:MAP_estimate_final}) using (\ref{eq:FIM_matrix}) we have
\begin{equation}
    \ln\big| \hat{\mathbf{F}}_\mathcal{I}\big | = \ln (N/2) -(k+2)\ln \hat{\sigma}^2_\mathcal{I} + \ln \left\lvert\mathbf{A}_\mathcal{I}^T\mathbf{A}_\mathcal{I}\right\rvert.\label{eq:ln_FIM}
\end{equation}
Here, the following property of the design matrix $\mathbf{A}$ is assumed \cite{Stoica_BIC_SNR,Schmidt_2011_Consistency_of_MDL}
\begin{equation}
    \lim_{N\to \infty} \left\{ N^{-1}(\mathbf{A}^T_\mathcal{I}\mathbf{A}_\mathcal{I})\right\} =\mathbf{M}_\mathcal{I}=\mathcal{O}(1),
    \label{eq:positive_definite}
\end{equation}
where $\mathbf{M}_\mathcal{I}$ is a $k\times k$ positive definite matrix and bounded as $N \to \infty$.
The assumption in (\ref{eq:positive_definite}) is true in many applications but not all (see \cite{djuric1998asymptotic} for more details). Using (\ref{eq:positive_definite}), it is possible to show that for large $N$
\begin{equation}
 \ln \big|\mathbf{A}_\mathcal{I}^T\mathbf{A}_\mathcal{I}\big |=\ln \bigg |N\cdot {N^{-1}(\mathbf{A}_\mathcal{I}^T\mathbf{A}_\mathcal{I})}\bigg| = k\ln N + \mathcal{O}(1).
\end{equation}
Furthermore, $\hat{\sigma}^2_\mathcal{I}$ is considered to be of $\mathcal{O}(1)$ as well since it does not grow with $N$.
As such, the $\mathcal{O}(1)$ term, $(k+2)\ln \hat{\sigma}^2_\mathcal{I}$ and $\ln (N/2)$ (a constant) are ignored from (\ref{eq:ln_FIM}). This leads to the final form of the BIC
\begin{equation}
    \text{BIC}(\mathcal{I}) = N\ln \hat{\sigma}^2_\mathcal{I} + k\ln N \label{eq:BIC}.
\end{equation}
BIC is consistent when $p$ is fixed and $N\to \infty$. However, it is inconsistent when $N$ is fixed and $\sigma^2 \to 0$ \cite{inconsistency_of_BIC_Kay_2011,GOHAIN-BIC-R} as well as when $p > N$ and $p$ grows exponentially with $N$ \cite{EBIC2008}.
\subsection{EBIC}
The authors in \cite{EBIC2008} proposed an extended version of the BIC, i.e., EBIC, to mitigate the drawbacks of BIC for large-$p$ small-$N$ scenarios. EBIC can be derived from the MAP estimator in (\ref{eq:MAP_estimate_final}), using the same assumptions as in BIC, except for the prior probability term $\Pr(\mathcal{H}_\mathcal{I})$. In EBIC, the idea of equiprobable models is discredited and instead a prior probability is assigned that is inversely proportional to the size of the model space. Thus, a model with dimension $k$ is assigned prior probability of $\Pr(\mathcal{H}_\mathcal{I}) \propto {p\choose k}^{-\gamma}$, where $0\leq \gamma \leq 1$ is a tuning parameter. Thus, the EBIC is
\begin{equation}
    \text{EBIC}(\mathcal{I}) =  N\ln \hat{\sigma}^2_\mathcal{I} + k\ln N + 2\gamma\ln {p\choose k}.
\end{equation}
When $\gamma = 0$, EBIC boils down to BIC (\ref{eq:BIC}).
Moreover, unlike BIC, EBIC is consistent in selecting the true model for $p\gg N$ cases where $p$ grows exponentially with $N$. However, it has been observed in \cite{EFIC2018} that EBIC is inconsistent when $N$ is fixed and $\sigma^2 \to 0$.

\subsection{EFIC}
To circumvent the shortcomings of EBIC in high-SNR cases, the authors in \cite{EFIC2018} proposed EFIC. In EFIC, the assumptions imposed on the sample FIM (\ref{eq:ln_FIM}) are removed and the entire structure is included as it is in the criterion except for the constant term $\ln (N/2)$. Some further simplifications are involved:
\begin{align}
    &N\ln \hat{\sigma}^2_\mathcal{I} = N \ln \big\lVert \mathbf{\Pi}^\perp_\mathcal{I}\mathbf{y} \big\rVert^2_2-N\ln N\label{eq:EFIC_simplify_1}\\
    &(k+2)\ln \hat{\sigma}^2_\mathcal{I} = (k+2)\left[ \ln\big\lVert \mathbf{\Pi}^\perp_\mathcal{I}\mathbf{y} \big\rVert^2_2-\ln N\right]\label{eq:EFIC_simplify_2}.
\end{align}
The $-N\ln N$ and $-2\ln N$ term of (\ref{eq:EFIC_simplify_1}) and (\ref{eq:EFIC_simplify_2}) respectively are independent of the model dimension $k$ and hence ignored. Similar to EBIC the prior probability term is assumed to be proportional to the model space, hence $\Pr (\mathcal{H}_\mathcal{I}) \propto  {p\choose k}^{-c} $, where $c>0$ is a tuning parameter. Furthermore, under the large-$p$ approximation and since $k\leq K\ll p$, the $\ln {p \choose k}$ term is approximated as
\begin{equation}
    \ln {p\choose k}=\sum_{i=0}^{k-1}\ln (p-i) -\ln (k!) \approx k\ln p.
\end{equation}
Hence, for large-$p$ case, we can set $ -2\ln p(\mathcal{H}_\mathcal{I}) \approx 2ck \ln p.$
Thus, the EFIC is given as
\begin{equation}
\begin{split}
    \text{EFIC}(\mathcal{I}) = N\ln \big\lVert \mathbf{\Pi}^\perp_\mathcal{I}\mathbf{y}    \big\rVert^2_2 + k\ln N 
     +\ln \big| \mathbf{A}^T_\mathcal{I}\mathbf{A}_\mathcal{I}\big|\\-(k+2)\ln \big\lVert \mathbf{\Pi}^\perp_\mathcal{I}\mathbf{y}    \big\rVert^2_2 + 2ck\ln p.
\end{split}
\end{equation}
EFIC is consistent in both large-$N$ and high-SNR scenarios \cite{EFIC2018}. However, EFIC suffers from a data scaling problem due to the inclusion of the data dependent penalty term and as such the performance of EFIC is not invariant to data scaling. See further in Section \ref{Sec:scaling-comparison}.

\section{Proposed criterion: EBIC-Robust (EBIC$_\text{R}$)}
In this section, we present the necessary steps for deriving EBIC$_\text{R}$. EBIC$_\text{R}$ can be seen as a natural extension of BIC$_\text{R}$ \cite{GOHAIN-BIC-R} for performing model selection in large-$p$ small-$N$ scenarios. Below, we provide a detailed derivation and establish the connection to BIC$_\text{R}$. A similar approach as in \cite{Stoica_BIC_SNR} is considered, but here we perform normalization of $\hat{\mathbf{F}}_\mathcal{I}$ under both large-$N$ and high-SNR assumption. It is possible to factorize the $\ln \big\lvert \hat{\mathbf{F}}_\mathcal{I}\big\rvert$ term in (\ref{eq:MAP_estimate_final}) in the following manner
\begin{align}
    \ln{\big|\hat{\mathbf{F}}_\mathcal{I}\big|} = & \ln \left[\big|\mathbf{L}\big|\left \lvert \mathbf{L}^{-1/2}\hat{\mathbf{F}}_\mathcal{I} \mathbf{L}^{-1/2}\right \rvert\right ]\nonumber\\
    = & \ln|\mathbf{L}| + \ln \underbrace{\Big | \mathbf{L}^{-1/2} \mathbf{\hat{F}}_\mathcal{I} \mathbf{L}^{-1/2}\Big |}_{\text{T}}.\label{eq:ln_det_FIM_EBIC_R}
\end{align}
The goal here is to choose a suitable $\mathbf{L}$ matrix that normalizes the sample FIM $\hat{\mathbf{F}}_\mathcal{I}$ such that the T term in (\ref{eq:ln_det_FIM_EBIC_R}) is $\mathcal{O}(1)$, i.e., in this case T should be bounded as $N \to \infty$ and/or $\sigma^2 \to 0$. To accomplish this objective, we choose the following $\mathbf{L}^{-1/2}$ matrix
\begin{equation}
    \mathbf{L}^{-1/2} = \begin{bmatrix}
    \sqrt{\frac{1}{N}}\sqrt{\frac{\hat{\sigma}_\mathcal{I}^2}{\hat{\sigma}_0^2}}\mathbf{I}_k & \mathbf{0}\\
    \mathbf{0} & \sqrt{\frac{1}{N}}\frac{\hat{\sigma}_\mathcal{I}^2}{\hat{\sigma}_0^2}
    \end{bmatrix},\label{eq:L_BIC_R}
\end{equation}
where $\hat{\sigma}^2_0 = \lVert \mathbf{y} \rVert^2_2/N$. The factor, $\hat{\sigma}^2_0$, is used in $\mathbf{L}^{-1/2}$ in order to neutralize the data scaling problem and is motivated by the fact that given (\ref{eq:positive_definite}), when the SNR is a constant, we have 
\begin{equation}
    \mathbb{E}[\hat{\sigma}^2_0] \to \text{const.} \quad \& \quad \text{Var}[\hat{\sigma}^2_0] \to 0 \label{eq:sigma2_0_limits}
\end{equation}
as $ N \to \infty$. Furthermore, from the considered generating model in (\ref{eq:LR_model_1}), when $N$ is fixed, (\ref{eq:sigma2_0_limits}) is also satisfied as $\sigma^2 \to 0$ (see Appendix \ref{appendix_sigma2_0_analysis} for details on $\hat{\sigma}^2_0$). Now using (\ref{eq:FIM_matrix}), (\ref{eq:L_BIC_R})  and the assumptions in (\ref{eq:positive_definite}), (\ref{eq:sigma2_0_limits}) it is possible to show that
\begin{equation}
    \Big | \mathbf{L}^{-1/2} \mathbf{\hat{F}}_\mathcal{I} \mathbf{L}^{-1/2}\Big | = \begin{vmatrix}
    {\frac{1}{\hat{\sigma}_0^2}}{\frac{\mathbf{A}^T_\mathcal{I}\mathbf{A}_\mathcal{I}}{N}} & \mathbf{0}\\
    \mathbf{0} & \frac{1}{2\hat{\sigma}_0^4} 
    \end{vmatrix}= \mathcal{O}(1),
    \label{eq:normalization_EBIC_R}
\end{equation}
and therefore may be discarded without much effect on the criterion. Furthermore, the $\ln \big \lvert \mathbf{L}\big \rvert$ term can be expanded as follows
\begin{align}
    \ln|\mathbf{L}| &=
    \ln \begin{vmatrix}
    N\left(\frac{\hat{\sigma}_0^2}{\hat{\sigma}^2_\mathcal{I}}\right)\mathbf{I}_k & \mathbf{0}\\
    \mathbf{0} & N\left (\frac{\hat{\sigma}^2_0}{\hat{\sigma}^2_\mathcal{I}}\right )^2
    \end{vmatrix}\nonumber\\
    &= (k+1)\ln N +(k+2)\ln  \left(\frac{\hat{\sigma}_0^2}{\hat{\sigma}_\mathcal{I}^2}\right ). \label{eq:ln_det_L_EBIC_R} 
\end{align}
Therefore, using (\ref{eq:normalization_EBIC_R}) and (\ref{eq:ln_det_L_EBIC_R}) we can rewrite (\ref{eq:ln_det_FIM_EBIC_R}) as
\begin{equation}
    \ln{\big|\hat{\mathbf{F}}_\mathcal{I}\big|}=k\ln N + (k+2)\ln  \left(\frac{\hat{\sigma}_0^2}{\hat{\sigma}_\mathcal{I}^2}\right) +\mathcal{O}(1) + \ln N. \label{eq:ln_FIM_EBIC_R_final}
\end{equation}
Next, for the model prior probability term $-2\ln \Pr(\mathcal{H}_\mathcal{I})$ in (\ref{eq:MAP_estimate_final}), a similar proposition is taken as in EBIC such that $\Pr(\mathcal{H}_\mathcal{I}) \propto  {p \choose k}^{-\zeta}$, where $\zeta \geq 0$ is a tuning parameter. For large-$p$, we follow a similar approach as in EFIC by employing the following approximation $\ln  {p \choose k} \approx k\ln p$ . This gives 
\begin{equation}
    -2\ln \Pr(\mathcal{H}_\mathcal{I}) = 2\zeta k \ln p + \text{const}.\label{eq:model-prior-ebic-r}
\end{equation}
Now, substituting (\ref{eq:ln_FIM_EBIC_R_final}), (\ref{eq:model-prior-ebic-r}) in  (\ref{eq:MAP_estimate_final}) and dropping the $\mathcal{O}(1)$, the $\ln N$ term (independent of $k$), the constant and the $p(\hat{\boldsymbol{\theta}}_\mathcal{I}|\mathcal{H}_\mathcal{I})$ term we arrive at the EBIC$_\text{R}$:
\begin{equation}
\begin{split}
    \text{EBIC}_\text{R}(\mathcal{I}) =&  N\ln \hat{\sigma}^2_\mathcal{I} + k\ln \left(\frac{N}{2\pi}\right)\\& + (k+2)\ln  \left(\frac{\hat{\sigma}_0^2}{\hat{\sigma}_\mathcal{I}^2}\right) + 2k\zeta \ln p.
     \label{eq:EBIC_R}
\end{split}
\end{equation}
The true model is estimated as
\begin{equation}
    \hat{\mathcal{S}}_{\text{EBIC}_\text{R}} = \underset{\mathcal{I}\in \mathcal{J} }{\arg \min} \big \{ \text{EBIC}_\text{R} (\mathcal{I})\big \},
\end{equation}
where $\mathcal{J}$ denotes the set of candidate models.

It can be observed from (\ref{eq:EBIC_R}) that the penalty of  EBIC$_\text{R}$ is a function of the number of measurements $N$, the ratio $(\hat{\sigma}_0^2/\hat{\sigma}_\mathcal{I}^2)$ and the parameter dimension $p$. Notice that the ratio $(\hat{\sigma}_0^2/\hat{\sigma}_\mathcal{I}^2)$ is always greater than $1$ and independent of the scaling of $\mathbf{y}$. Furthermore, when $\mathcal{S}\not\subset \mathcal{I}$, the ratio $(\hat{\sigma}_0^2/\hat{\sigma}_\mathcal{I}^2) \approx \mathcal{O}(1)$ and for $\mathcal{S}\subset \mathcal{I}$ we have $(\hat{\sigma}_0^2/\hat{\sigma}_\mathcal{I}^2) \approx \mathcal{O}(\text{SNR}+1)$. Hence, the behaviour of the penalty can be summarized as follows: (i) For fixed $p$ and SNR, as $N \to \infty $ the penalty grows as $\mathcal{O}(\ln N)$; (ii) If $N$ and $p$ are constant, as SNR $\to \infty$, the penalty grows as $\mathcal{O}\left(\ln (\text{SNR}+1)\right)$ for all $\mathcal{I}\supset \mathcal{S}$;
(iii) when SNR is a constant and given that $p$ grows with $N$, then as $N\to\infty$ the penalty grows as $\mathcal{O}(\ln N) +\mathcal{O}(\ln p)$.


\subsection{Scaling Robustness as Compared to EFIC}\label{Sec:scaling-comparison}
In this section, we elaborately discuss the data scaling problem. Ideally, any model selection criterion should be invariant to data scaling, which means that if $\mathbf{y}$ is scaled by any arbitrary constant $C>0$, the equivalent penalty for each of the models $\mathcal{I}$ should not change. This property is necessary because otherwise the behaviour of the model selection criterion will be unreliable and may suffer from overfitting or underfitting issues when the data is scaled. As mentioned before, the penalty of EFIC is not invariant to data scaling. This can be observed from the following analysis. Let $\Delta = {card}(\mathcal{I})-{card}(\mathcal{S})$. Now, consider the difference assuming $\mathcal{I} \neq \mathcal{S}$
\begin{align}
    &\text{EFIC}(\mathcal{I}) - \text{EFIC}(\mathcal{S}) \nonumber\\
    &= 
    \ (N-2)\ln \frac{\big\lVert\boldsymbol{\Pi}^\perp_\mathcal{I}\mathbf{y}\big\rVert^2_2}{\big\lVert\boldsymbol{\Pi}^\perp_\mathcal{S}\mathbf{y}\big\rVert^2_2} + \ln\frac{\big| \mathbf{A}^T_\mathcal{I}\mathbf{A}_\mathcal{I}\big|}{\big | \mathbf{A}^T_\mathcal{S}\mathbf{A}_\mathcal{S}\big|} - k\ln\big\lVert\boldsymbol{\Pi}^\perp_\mathcal{I}\mathbf{y}\big\rVert^2_2\nonumber\\
    &\ \ +k_0\ln \big\lVert\boldsymbol{\Pi}^\perp_\mathcal{S}\mathbf{y}\big\rVert^2_2 +\Delta \left(\ln N + 2 c \ln p\right)=D_\text{EFIC} \text{ (say)}.
    \label{eq:EFIC_diff_Unscaled}
\end{align}
Ideally, for correct model selection, $D_{\text{EFIC}}>0$ for all $ \mathcal{I}\neq \mathcal{S}$. Now, if we scale the data $\mathbf{y}$ by a constant $C>0$, the data dependent term becomes $\ln \lVert \boldsymbol{\Pi}^\perp_\mathcal{I} C\mathbf{y}\rVert^2_2 = \ln C^2 + \ln\lVert
\boldsymbol{\Pi}^\perp_\mathcal{I}\mathbf{y} \rVert^2_2$ and the difference becomes
\begin{equation}
    \text{EFIC}(\mathcal{I}) - \text{EFIC}(\mathcal{S}) = D_\text{EFIC} - \Delta \ln C^2. \label{eq:EFIC_diff_scaled}
\end{equation}
It is evident that (\ref{eq:EFIC_diff_Unscaled}) and (\ref{eq:EFIC_diff_scaled}) are unequal and the difference after scaling contains an additional term $-\Delta \ln C^2$. This implies that scaling the data changes the EFIC score difference between any arbitrary model $\mathcal{I}$ and the true model $\mathcal{S}$. Hence, depending on the $C$ value ($C<1$ or $C\geq 1$) and $\Delta>0$ or $\Delta<0$, the difference in (\ref{eq:EFIC_diff_scaled}) may become negative leading to a false model selection. Thus, EFIC is not invariant to data scaling. On the contrary, consider the difference for EBIC$_\text{R}$,
\begin{align}
    &\text{EBIC}_\text{R}(\mathcal{I}) - \text{EBIC}_\text{R}(\mathcal{S})\nonumber\\
    & = (N-2)\ln \left( \frac{\hat{\sigma}^2_\mathcal{I}}{\hat{\sigma}^2_\mathcal{S}}\right)-k\ln \hat{\sigma}^2_\mathcal{I} + k_0\ln \hat{\sigma}^2_\mathcal{S}  
    +\Delta \ln \hat{\sigma}^2_0 \nonumber\\
    &\ \ + \Delta\left(\ln (N/2\pi) +2\zeta \ln p\right)= D_{\text{EBIC}_\text{R}} \text{(say)}
    \label{eq:EBIC-R_diff_scaled}
\end{align}
Now, scaling $\mathbf{y}$ by $C$, scales the noise variance estimates $\hat{\sigma}^2_\mathcal{I}$, $\hat{\sigma}^2_\mathcal{S}$ and $\hat{\sigma}^2_0$ by $C^2$, however, the difference remains the same, i.e., $D_{\text{EBIC}_\text{R}}$.
This is because in this case the $-\Delta \ln C^2$ term is cancelled by $+\Delta \ln C^2$ generated by $\Delta \ln \hat{\sigma}^2_0$. Hence, EBIC$_\text{R}$ is invariant to data scaling, which is a desired property of any model selection criterion.

\section{Consistency of EBIC$_\text{R}$}
In this section, we provide the necessary proofs to show that EBIC$_\text{R}$ is a consistent criterion. Generally speaking, a model selection criterion with $\hat{\mathcal{S}}$ as its estimate of the true model $\mathcal{S}$ is consistent if it satisfies the following conditions \cite{EFIC2018} 
\begin{equation}
    \lim_{\sigma^2 \to 0}\Pr \{ \hat{\mathcal{S}} = \mathcal{S}\} =1 \quad \& \quad \lim_{N \to \infty}\Pr \{ \hat{\mathcal{S}} = \mathcal{S}\} =1.
\end{equation}
Let us define the set of all overfitted models of dimension $k$ as $\mathcal{I}_o^k = \left\{ \mathcal{I} : {card}(\mathcal{I}) = k, \mathcal{S} \subset \mathcal{I} \right\}$ and the set of all misfitted models of dimension $k$ as $\mathcal{I}^k_m = \left\{ \mathcal{I} : {card}(\mathcal{I}) = k, \mathcal{S} \not\subset \mathcal{I} \right\}$. Furthermore, let $\mathbb{O}$ denote the set of all $\mathcal{I}^k_o$ for $k=k_0+1,\ldots,K$, and let $\mathbb{M}$ denote the set of all $\mathcal{I}^k_m$ for $k=1,\ldots,K$, i.e.,
\begin{equation}
    \mathbb{O}=\bigcup_{k=k_0+1}^{K} \mathcal{I}^k_o \quad \text{and} \quad 
    \mathbb{M}=\bigcup_{k=1}^{K} \mathcal{I}^k_m\ ,
\end{equation}
where $K$ is some upper bound for $k_0$ and $k_0\leq K\ll N$. In practice, EBIC$_\text{R}$ picks the true model $\mathcal{S}$, if the following conditions are satisfied:
\begin{align}
    &\mathcal{C}_1:\text{EBIC}_\text{R}(\mathcal{S})<  \text{EBIC}_\text{R}(\mathcal{I})  \quad \forall\ \mathcal{I} \in \mathbb{O} \label{eq:C1}\\
    &\mathcal{C}_2:\text{EBIC}_\text{R}(\mathcal{S})<  \text{EBIC}_\text{R}(\mathcal{I})  \quad \forall\ \mathcal{I} \in \mathbb{M}.\label{eq:C2}
\end{align}
\subsection{Asymptotic Identifiability of the Model}
In general, the model is identifiable if no model of comparable size other than the true submodel can predict the noise free response almost equally well \cite{EBIC2008}. In the context of linear regression, this is equivalent to say $\mathbf{y} = \mathbf{A}_\mathcal{S}\mathbf{x}_\mathcal{S} \neq \mathbf{A}_\mathcal{I}\mathbf{x}_\mathcal{I}$ for $\big \{ \mathcal{I} : {card}(\mathcal{I}) \leq {card}(\mathcal{S}), \mathcal{I} \neq \mathcal{S}\big\}$. The identifiability of the true model in the high-dimensional linear regression setup is uniformly maintained if the minimal eigenvalue of all restricted sub-matrices, $\mathbf{A}^T_\mathcal{I}\mathbf{A}_\mathcal{I}$ for $\{\mathcal{I}:{card}(\mathcal{I}) \leq 2K\}$, is bounded away from zero \cite{EFIC2018}. 
A sufficient assumption on the design matrix $\mathbf{A}$ to prove the consistency of EBIC$_\text{R}$ is the sparse Riesz condition \cite{zhang2008sparsity}:
\begin{equation}
     \lim\limits_{N\to \infty}\left\{N^{-1}\left(\mathbf{A}_\mathcal{I}^T\mathbf{A}_\mathcal{I}\right)\right\} = \mathbf{M}_\mathcal{I},\quad  \forall\  card(\mathcal{I})\leq 2K,
    \label{eq:identifiability-condition-on-A}
\end{equation}
where $\mathbf{M}_\mathcal{I}$ denotes a bounded positive definite matrix.

\subsection{Consistency as $\sigma^2 \to 0$ or $\text{SNR} \to \infty $ for fixed $N$}
In this subsection, we examine whether EBIC$_\text{R}$ selects the true model $\mathcal{S}$ as $\sigma^2$ goes vanishingly small (or equivalently SNR$\to \infty$) under the assumption that $N$ is fixed. We formulate this into a theorem as follows:

\begin{theorem}\label{theorem-high-SNR}
Assume that $N$ and $p$ are fixed and the matrix $\mathbf{A}$ satisfies the condition given by (\ref{eq:identifiability-condition-on-A}). If $K\geq k_0$, then $\Pr \left \{\text{EBIC}_\text{R} (\mathcal{S})< \text{EBIC}_\text{R} (\mathcal{I})\right\} \to 1$ as $\sigma^2 \to 0$ for all $\mathcal{I}\neq \mathcal{S}$ and $card(\mathcal{I}) = 1,\ldots,K$.
\end{theorem}

\textit{Proof.} The proof consists of two parts. In part (a) we show that the probability of overfitting $(\mathcal{S}\subset \hat{\mathcal{S}}_{\textrm{EBIC}_\textrm{R}} )$ tends to $0$ as $\sigma^2 \to 0$, which in this case is equivalent to showing $\lim_{\sigma^2 \to 0}\Pr(\mathcal{C}_1) = 1$, cf. (\ref{eq:C1}). In part (b) we show that the probability of misfitting $(\mathcal{S}\not\subset \hat{\mathcal{S}}_{\textrm{EBIC}_\textrm{R}} )$ also tends to $0$ as $\sigma^2 \to 0$, which is equivalent to $\lim_{\sigma^2 \to 0}\Pr(\mathcal{C}_2) = 1$, cf. (\ref{eq:C2}).

$(a)$ \textit{Over-fitting case} $(\mathcal{S}\subset \hat{\mathcal{S}}_{\textrm{EBIC}_\textrm{R}} )$:
Consider the set of overfitted subsets having cardinality $k$, which we have denoted as $\mathcal{I}^k_o$. Let $\mathcal{I}_j$ denote the $j$th subset in the set $\mathcal{I}^k_o$. The total number of subsets in $\mathcal{I}^k_o$ is ${p-k_0}\choose{\Delta}$ where $\Delta=k-k_0$ .  For any overfitted subset $\mathcal{I}_j \in \mathcal{I}^k_o$, consider the following inequality
\begin{equation}
    \text{EBIC}_\text{R}(\mathcal{S})<  \text{EBIC}_\text{R}(\mathcal{I}_j),  \quad \mathcal{I}_j \in \mathcal{I}^k_o, \label{eq:inequality_overfit_sigma2}
\end{equation}
where $j = 1,\ldots, {{p-k_0}\choose{\Delta}}$.
Using the relation $p = N^d$ and after some straightforward rearrangement of (\ref{eq:inequality_overfit_sigma2}) we get
\begin{align}
    (N-k_0-2)\ln \hat{\sigma}^2_\mathcal{S}&-(N-k-2)\ln \hat{\sigma}^2_{\mathcal{I}_j}\nonumber\\
      -\Delta (1+2\zeta d)&\ln N- \Delta \ln \hat{\sigma}^2_0 + \Delta \ln 2\pi < 0. \label{eq:overfit_sigma_1}
\end{align}
Let us define a random variable $X_{\mathcal{I}_j} = {\hat{\sigma}^2_{\mathcal{I}_j}}/{\sigma^2}$, then
\begin{equation}
    N \cdot X_{\mathcal{I}_j} \sim \chi^2_{N-k}, \quad \forall\  \mathcal{I}_j \in \mathcal{I}^k_o. \label{eq:X_RV}
\end{equation}
This implies that the variables $X_{\mathcal{I}_j}$ are independent of $\sigma^2$. Now, we can express
\begin{equation}
(N-k-2)\ln \hat{\sigma}^2_{\mathcal{I}_j}=\ln X_{\mathcal{I}_j}^{N-k-2} +(N-k-2)\ln \sigma^2, \label{eq:overfit_sigma_split_1}
\end{equation}
and similarly by defining $X_\mathcal{S} = \hat{\sigma}^2_\mathcal{S}/\sigma^2$ we get
\begin{equation}
(N-k_0-2)\ln \hat{\sigma}^2_\mathcal{S} = \ln X_\mathcal{S}^{N-k_0-2} +(N-k_0-2)\ln \sigma^2. \label{eq:overfit_sigma_split_2}
\end{equation}
Using (\ref{eq:overfit_sigma_split_1}) and (\ref{eq:overfit_sigma_split_2}) in (\ref{eq:overfit_sigma_1}) and after exponentiation we get
\begin{align}
    \left(\frac{X^{N-k_0-2}_\mathcal{S}}{X^{N-k-2}_{\mathcal{I}_j}}\right)  \left( \frac{1}{N} \right)^{\Delta(1+2\zeta d)} \left( \frac{2\pi}{\hat{\sigma}^2_0}\right)^\Delta
     <  \left( \frac{1}{\sigma^2}\right)^\Delta. \label{eq:overfit_sigma_2} 
\end{align}
Let $E^k_{\mathcal{I}_j}$ denote the entire left hand-side and let $\eta_k$ denote the right-hand side of the inequality in (\ref{eq:overfit_sigma_2}). 
Let $\mathcal{I}^*\in \mathcal{I}^k_o$ denote the subset that produces the maximum value of $E_{\mathcal{I}_j}^k$ among all such subsets $\mathcal{I}_j \in \mathcal{I}^k_o$ . Then, let us denote
\begin{equation}
    E^k_{\mathcal{I}^*}= \underset{\mathcal{I}_j \in \mathcal{I}^k_o}{\max} \left\{ E^k_{\mathcal{I}_j} \right\} , \quad j = 1,2,\ldots, {{p-k_0}\choose{\Delta}}.
\end{equation}
The condition $\mathcal{C}_1$ in (\ref{eq:C1}) is satisfied as $\sigma^2 \to 0$ under the event $E^k_{\mathcal{I}^*} < \eta_k,$ for all $k=k_0+1,\ldots, K.$
Now, we can express the probability that $E^k_{\mathcal{I}^*} < \eta_k$ as follows
\begin{align}
    \Pr \left(E^k_{\mathcal{I}^*}<\eta_k \right) &= \Pr\left \{\bigcap_{j=1}^{{p-k_0}\choose{\Delta}} \left(E^k_{\mathcal{I}_j}<\eta_k\right)\right\} \nonumber\\ 
    &= 1- \Pr\left \{\bigcup_{j=1}^{{p-k_0}\choose{\Delta}} \left(E^k_{\mathcal{I}_j}>\eta_k\right)\right\}\nonumber   \\
    &\geq 1 - {{p-k_0}\choose{\Delta}} \Pr \left(E^k_{\mathcal{I}_j}>\eta_k\right) \nonumber\\
    \implies \Pr \left(E^k_{\mathcal{I}^*}>\eta_k \right)&\leq {{p-k_0}\choose{\Delta}} \Pr \left(E^k_{\mathcal{I}_j}>\eta_k\right),\label{eq:overfit_sigma_Pr_Ek_Istar_bound}
\end{align}
where the inequality follows from the union bound. Now consider the following probability $\Pr \left\{E^k_{\mathcal{I}_j}>\eta_k\right\}$ for any arbitrary subset $\mathcal{I}_j\in \mathcal{I}^k_o$, which can be expressed as
\begin{align}
    &\Pr \left \{\left(\frac{X^{N-k_0-2}_\mathcal{S}}{X^{N-k-2}_{\mathcal{I}_j}}\right)  \left( \frac{1}{N} \right)^{\Delta(1+2\zeta d)} \left( \frac{2\pi}{\hat{\sigma}^2_0}\right)^\Delta 
     >  \left( \frac{1}{\sigma^2}\right)^\Delta\right \}.\label{eq:overfit_sigma_EkIj_greater_eta_k}
\end{align}
Let $W = {X^{N-k_0-2}_\mathcal{S}}/{X^{N-k-2}_{\mathcal{I}_j}}$. Notice that the random variable $W$ is independent of the noise variance $\sigma^2$ and since $N$ is fixed $W$ is bounded as $\sigma^2 \to 0$. Furthermore, $\lim\limits_{\sigma^2 \to 0}\hat{\sigma}^2_0 = c$ (see Appendix \ref{appendix_sigma2_0_analysis}) and the right-hand side of the inequality in (\ref{eq:overfit_sigma_EkIj_greater_eta_k}) grows unbounded as $\sigma^2\to 0$. Thus, we have
\begin{align}
    \lim_{\sigma^2 \to 0}\Pr \left \{ E^k_{\mathcal{I}_j}>\eta_k\right\} = 0.\label{eq:overfit_sigma_Ek_Istar_greater_etak}
\end{align}
Therefore, using (\ref{eq:overfit_sigma_Pr_Ek_Istar_bound}) and the result in (\ref{eq:overfit_sigma_Ek_Istar_greater_etak}), we have
\begin{align}
 \lim_{\sigma^2 \to 0}\Pr\left(E^k_{\mathcal{I}^*}>\eta_k\right) = 0, \qquad \forall\ k = k_0+1,\ldots, K .\label{eq:overfit_sigma_Ek_Istar_greatet_eta}
\end{align}
Finally, using the union bound, and the result in (\ref{eq:overfit_sigma_Ek_Istar_greatet_eta}), we get
\begin{align}
    \Pr \left \{ \mathcal{C}_1\right \} = &\Pr \left \{ \bigcap_{k=k_0+1}^{K}E^k_{\mathcal{I}^*}<\eta_k\right\}\nonumber\\
    \geq & 1-\sum_{k=k_0+1}^{K}\Pr \left\{ E^k_{\mathcal{I}^*}>\eta_k \right\} \to 1,\label{eq:overfit_sigma_C1}
\end{align}
as $\sigma^2 \to 0$.

$(b)$ \textit{Misfitting case} $(\mathcal{S}\not \subset \hat{\mathcal{S}}_{\textrm{EBIC}_\textrm{R}} )$:
Let $\mathcal{I}_j$ be any arbitrary $j$th subset belonging to the set of misfitted subsets of dimension $k$, i.e., $\mathcal{I}^k_m$. 
We consider the following inequality
\begin{equation}
    \text{EBIC}_\text{R}(\mathcal{S})<  \text{EBIC}_\text{R}(\mathcal{I}_j), \quad \mathcal{I}_j \in \mathcal{I}^k_m,  \label{eq:proof-EBIC-R<EBIC-I}
\end{equation}
where $j =1, \ldots,t$.  Here, $t$ denotes the total number of subsets in the set $\mathcal{I}^k_m$ and $t={{p}\choose{k}}$ if $k<k_0$, otherwise $t = {{p}\choose{k}}-{{p-k_0}\choose{\Delta}}$ if $k\geq k_0$, where $\Delta = k-k_0$. Denoting $X_\mathcal{S} = \hat{\sigma}^2_\mathcal{S}/\sigma^2$, rearranging and applying exponentiation we can express (\ref{eq:proof-EBIC-R<EBIC-I}) as
\begin{align}
 \left(\frac{X^{N-k_0-2}_\mathcal{S}}{(\hat{\sigma}^2_{\mathcal{I}_j})^{N-k-2}}\right)  \left( \frac{1}{N} \right)^{\Delta(1+2\zeta d)} \left( \frac{2\pi}{\hat{\sigma}^2_0}\right)^\Delta 
     <  \left( \frac{1}{\sigma^2}\right)^{N-k_0-2}. \label{eq:misfit_sigma_1}
\end{align}
Similar to the overfitting case, let $E^k_{\mathcal{I}_j}$ denote the entire left-hand side and $\eta$ the right-hand side of (\ref{eq:misfit_sigma_1}). Also, let $E^k_{\mathcal{I}^*} = \underset{\mathcal{I}_j \in \mathcal{I}_m^k}{\max} \left\{ E^k_{\mathcal{I}_j}\right\}$ for $j=1,\ldots,t$, where $\mathcal{I}^*$ is the subset that leads to the maximum value of $E^k_{\mathcal{I}_j}$ among all such subsets of dimension $k$. The condition $\mathcal{C}_2$ in (\ref{eq:C2}) is satisfied as $\sigma^2 \to 0$ under the event $E^k_{\mathcal{I}^*}<\eta, $ for all $   k = 1,\ldots,K.$ Now, we can express the probability that $E^k_{\mathcal{I}^*}< \eta$ as
\begin{align}
    \Pr \left(E^k_{\mathcal{I}^*}<\eta \right) =& \Pr\left \{\bigcap_{j=1}^{t} \left(E^k_{\mathcal{I}_j}<\eta\right)\right\}\nonumber\\
\implies \Pr \left(E^k_{\mathcal{I}^*}>\eta \right) \leq &  t \Pr \left(E^k_{\mathcal{I}_j}>\eta\right),\label{eq:misfit_sigma_EkIstar_greater_eta}
\end{align}
where the inequality follows from the union bound. Now consider the following probability for any arbitrary subset $\mathcal{I}_j \in \mathcal{I}^k_m$
\begin{equation}
    \begin{split}
     \Pr \left (E^k_{\mathcal{I}_j} > \eta\right)
    = \Pr \bigg \{ \left(\frac{X^{N-k_0-2}_\mathcal{S}}{(\hat{\sigma}^2_{\mathcal{I}_j})^{N-k-2}}\right)  \left( \frac{1}{N} \right)^{\Delta(1+2\zeta d)}\\
    \times \left( \frac{2\pi}{\hat{\sigma}^2_0}\right)^\Delta 
     >  \left( \frac{1}{\sigma^2}\right)^{N-k_0-2} \bigg\}.\label{eq:misfit_sigma_EkIj_greater_eta}
    \end{split}
\end{equation}
Here, $X^{N-k_0-2}_\mathcal{S}$ is independent of $\sigma^2$ and $N$ is fixed, therefore $X^{N-k_0-2}_\mathcal{S}$ is bounded as $\sigma^2 \to 0$. Also $\hat{\sigma}^2_{\mathcal{I}_j} \to \lVert \boldsymbol{\Pi}^\perp_{\mathcal{I}_j}\mathbf{A}_\mathcal{S}\mathbf{x}_\mathcal{S}\rVert^2_2/N$ in probability as $\sigma^2\to 0$ and since we are in the misfitting scenario, from Lemma \ref{lemma-identifiability} in Appendix \ref{appendix_lemmas-2-to-4} we have $ \lVert \boldsymbol{\Pi}^\perp_{\mathcal{I}_j}\mathbf{A}_\mathcal{S}\mathbf{x}_\mathcal{S}\rVert^2_2/N > 0$. Furthermore, $\lim\limits_{\sigma^2 \to 0}\hat{\sigma}^2_0 = \text{const.}$ (see Appendix \ref{appendix_sigma2_0_analysis}) and the right-hand side of the inequality in  (\ref{eq:misfit_sigma_EkIj_greater_eta}) grows unbounded as $\sigma^2 \to 0$. Hence,
\begin{align}
    \lim_{\sigma^2 \to 0}\Pr \left \{ E^k_{\mathcal{I}_j} > \eta\right\}
    = 0 \label{eq:misfit_sigma_lim_EkIj_greater_eta}.
\end{align}
Using (\ref{eq:misfit_sigma_EkIstar_greater_eta}) and the result in (\ref{eq:misfit_sigma_lim_EkIj_greater_eta}) we get
\begin{align}
    \lim_{\sigma^2 \to 0}\Pr \left \{ E^k_{\mathcal{I}^*} > \eta\right\} = 0, \quad \forall \ k=1,\ldots, K.\label{eq:misfit_sigma_Ek_Istar_greatet_eta}
\end{align}
Finally, using the union bound and the result in (\ref{eq:misfit_sigma_Ek_Istar_greatet_eta}), we get
\begin{equation}
    \Pr \left \{ \mathcal{C}_2\right \}  \geq 1-\sum_{k=1}^{K}\Pr \left\{ E^k_{\mathcal{I}^*}>\eta \right\} \to 1 \quad \text{as} \quad \sigma^2 \to 0. \label{eq:misfit_sigma_C2}
\end{equation}
From (\ref{eq:overfit_sigma_C1}) and (\ref{eq:misfit_sigma_C2}) we can conclude that EBIC$_\text{R}$ is consistent as $\sigma^2 \to 0$, which proves Theorem 1.

\subsection{Consistency as $N \to \infty$ when $\sigma^2$ is fixed $(0<\sigma^2<\infty)$}
In this section, we prove the consistency of EBIC$_\text{R}$ as the sample size $N \to \infty$ given that $\sigma^2$ is fixed and under the setting $p = N^d$ for some $d>0$. This is a common setting in the model selection literature (see, e.g., \cite{EBIC2008,EFIC2018,meinshausen2006high}). This leads to the following theorem.

\begin{theorem}\label{theorem-large-N}
Assume that $p=N^d$ for some constant $d>0$, the SNR is fixed and the matrix $\mathbf{A}$ satisfies  (\ref{eq:identifiability-condition-on-A}). If $K\geq k_0$, then $\Pr \left\{ \text{EBIC}_\text{R} (\mathcal{S})< \text{EBIC}_\text{R} (\mathcal{I}) \right\} \to 1$ as $N \to \infty$ for all $\mathcal{I}\neq \mathcal{S}$ and $card(\mathcal{I}) = 1,\ldots,K$ under the condition $\zeta > 1 -1/2d$.
\end{theorem}

\textit{Proof.} As in the previous section, we have two parts of the proof. Part $(a)$ is the overfitting case where we show that $\Pr(\mathcal{C}_1) \to 1$ as $N \to \infty$ and part $(b)$ is the misfitting case where we show that $\Pr(\mathcal{C}_2) \to 1$ as $N \to \infty$.\\
$(a)$ \textit{Overfitting case} $(\mathcal{S} \subset \hat{\mathcal{S}}_{\text{EBIC}_\text{R}})$:
Let $\mathcal{I}_j \in \mathcal{I}^k_o$ be any overfitted subset of dimension $k$. 
Consider the following inequality
\begin{equation}
    \text{EBIC}_\text{R}(\mathcal{I}_j)>\text{EBIC}_\text{R}(\mathcal{S}), \quad \mathcal{I}_j \in \mathcal{I}^k_o.\label{eq:EFIC_S_less_EFIC_I_overfit_N_0}
\end{equation}
Denoting $\Delta = k-k_0$ and rearranging (\ref{eq:EFIC_S_less_EFIC_I_overfit_N_0}) we get
\begin{align}
    {\left(N-k-2\right)}\ln\left(\frac{\hat{\sigma}^2_{\mathcal{I}_j}}{\hat{\sigma}^2_{\mathcal{S}}}\right) + \Delta(1+2\zeta d) \ln N\nonumber\\
    +\Delta \ln \left(\frac{\hat{\sigma}^2_0}{\hat{\sigma}^2_{\mathcal{S}}}\right)
    - \Delta \ln 2\pi >0.
    \label{eq:overfitting_N_inequality}
\end{align}
Let $E^k_{\mathcal{I}_j}$ denote the entire left side of the inequality (\ref{eq:overfitting_N_inequality}) and  $\mathcal{I}^*$ denote the subset that leads to the minimum value of $E^k_{\mathcal{I}_j}$ among all such subsets of dimension $k$. Hence,
 \begin{equation}
     E^k_{\mathcal{I}^*} = \underset{\mathcal{I}_j \in \mathcal{I}^k_o}{\min} \left\{ E^k_{\mathcal{I}_j} \right\}, \quad j = 1,2,\ldots , {{p-k_0}\choose{\Delta}}.
 \end{equation}
The condition $\mathcal{C}_1$ in (\ref{eq:C1}) is satisfied as $N \to \infty$ under the event $E^k_{\mathcal{I}^*} > 0,$ for all $ k=k_0+1,\ldots,K$. Expanding the ratio we have
\begin{align}
    \ln \left(\frac{\hat{\sigma}^2_{\mathcal{I}_j} }{\hat{\sigma}^2_{\mathcal{S}}}\right)
    &=\ln  \left(\frac{\mathbf{e}^T\boldsymbol{\Pi}^\perp_{\mathcal{I}_j}\mathbf{e}}{\mathbf{e}^T\boldsymbol{\Pi}^\perp_\mathcal{S}\mathbf{e}} \right)\nonumber\\
    &= \ln \left[\frac{\mathbf{e}^T \left(\mathbf{I}-\boldsymbol{\Pi}_{\mathcal{I}_j}+\boldsymbol{\Pi}_\mathcal{S}-\boldsymbol{\Pi}_\mathcal{S}\right) \mathbf{e}}{\mathbf{e}^T\boldsymbol{\Pi}^\perp_\mathcal{S}\mathbf{e}}\right]\nonumber\\
    &= \ln \left(\frac{\mathbf{e}^T\boldsymbol{\Pi}^\perp_\mathcal{S}\mathbf{e}- \mathbf{e}^T\boldsymbol{\Pi}_{{\mathcal{I}_j}\setminus\mathcal{S}}\mathbf{e}}{\mathbf{e}^T\boldsymbol{\Pi}^\perp_\mathcal{S}\mathbf{e}}\right)\nonumber\\
    &= \ln \left(1-\frac{\mathbf{e}^T\boldsymbol{\Pi}_{{\mathcal{I}_j}\setminus\mathcal{S}}}{\mathbf{e}^T\boldsymbol{\Pi}^\perp_\mathcal{S}\mathbf{e}}\right),
\end{align}
where $\boldsymbol{\Pi}_{{\mathcal{I}_j}\setminus\mathcal{S}} = \boldsymbol{\Pi}_{\mathcal{I}_j}-\boldsymbol{\Pi}_\mathcal{S}$. Now we can write
\begin{align}
    &\underset{1\leq j\leq T}{\min} \Bigg \{(N-k-2)\ln \left(\frac{\hat{\sigma}^2_{\mathcal{I}_j} }{\hat{\sigma}^2_{\mathcal{S}}}\right)\Bigg \}=\nonumber\\ 
    &\ \ (N-k-2)\ln \left[1- \frac{\underset{1\leq j\leq T}{\max}\left\{\left(\mathbf{e}^T\boldsymbol{\Pi}_{{\mathcal{I}_j}\setminus\mathcal{S}}\mathbf{e}\right)/\sigma^2\right\}}{\left(\mathbf{e}^T\boldsymbol{\Pi}^\perp_\mathcal{S}\mathbf{e}\right)/\sigma^2} \right],\label{eq:min=max-overfit-N}
\end{align}
where $T = {p-k_0\choose \Delta}$. Now the term, $(\mathbf{e}^T\boldsymbol{\Pi}_{{\mathcal{I}_j}\setminus\mathcal{S}}\mathbf{e})/\sigma^2 \sim \chi^2_{\Delta}$ (see Appendix \ref{appendix_sigma2_I_analysis}). Then from Lemma \ref{lemma-Chi2-bound} in Appendix \ref{appendix_lemmas-2-to-4} we have the following upper bound
\begin{equation}
    \underset{1\leq j\leq T}{\max} \Big\{(\mathbf{e}^T\boldsymbol{\Pi}_{{\mathcal{I}_j}\setminus\mathcal{S}}\mathbf{e})/\sigma^2\Big\}  \leq \Delta + 2\sqrt{\Delta\psi \ln T} +2\psi \ln T,
\end{equation}
with probability approaching one as $N\to \infty$ if $\psi >1$. Now, for sufficiently large $p=N^d$ we can write $\ln T = \ln {p-k_0 \choose \Delta}\approx \Delta d \ln N$. This gives
\begin{align}
    &\small\underset{1\leq j\leq T}{\max} \Big\{(\mathbf{e}^T\boldsymbol{\Pi}_{{\mathcal{I}_j}\setminus\mathcal{S}}\mathbf{e})/\sigma^2\Big\}  \leq  \Delta + 2\Delta\sqrt{\psi d  \ln N} + 2\psi \Delta d\ln N \nonumber\\
    &\qquad \qquad \qquad =  2\psi \Delta d\ln N\left( 1+ \frac{1}{\sqrt{\psi d\ln N}} +\frac{1}{2\psi d\ln N} \right) \nonumber\\
    &\qquad \qquad \qquad \approx  2\psi \Delta d\ln N, \label{eq:max-bound-chi-ovefit-N}
\end{align}
as $N$ grows large. Furthermore, the term in the denominator in (\ref{eq:min=max-overfit-N}), $(\mathbf{e}^T\boldsymbol{\Pi}^\perp_\mathcal{S}\mathbf{e})/\sigma^2 \sim \chi^2_{N-k_0}$ and based on the law of large numbers tends to $N-k_0 \approx N$. Therefore, using (\ref{eq:max-bound-chi-ovefit-N}) in (\ref{eq:min=max-overfit-N}) and $(N-k-2) \approx N$ under the large-$N$ approximation we get
\begin{align}
    \underset{1\leq j\leq T}{\min} \Bigg \{N\ln \left(\frac{\hat{\sigma}^2_{\mathcal{I}_j} }{\hat{\sigma}^2_{\mathcal{S}}}\right)\Bigg \} &\geq N\ln \left(1- \frac{2\Delta \psi d \ln N}{N}\right) \nonumber\\
    &\approx -2\Delta \psi d \ln N,
\end{align}
where the last approximation follows by linearization of the logarithm for small $2\Delta \psi d \ln N/N$ value. Thus, we can write
\begin{align}
    E^k_{\mathcal{I}^*} &\geq -2\Delta \psi d \ln N + \Delta(1+2\zeta d) \ln N +\Delta \ln \left(\frac{\hat{\sigma}^2_0}{\hat{\sigma}^2_{\mathcal{S}}}\right)\nonumber\\ 
    & \hspace{5cm}- \Delta \ln 2\pi \nonumber\\
    & = \Delta \left( 1+2\zeta d -2\psi d \right)\ln N+\Delta \ln \left(\frac{\hat{\sigma}^2_0}{\hat{\sigma}^2_{\mathcal{S}}}\right) - \Delta \ln 2\pi.
\end{align}
Since $\lim_{N \to \infty} \hat{\sigma}^2_{0} = \text{const.} >0$ (see Appendix \ref{appendix_sigma2_0_analysis}) and $\lim_{N \to \infty} \hat{\sigma}^2_\mathcal{S} = \sigma^2$ (see Appendix \ref{appendix_sigma2_I_analysis}), $E^k_{\mathcal{I}^*} \to \infty$ as $N\to \infty$ for all $k=k_0+1, \ldots, K$  under the condition $1+2\zeta d-2 \psi d >0$ for any $\psi>1$. Hence, the lower bound on $\zeta$ becomes
\begin{equation}
    \boxed{\zeta > 1 - \frac{1}{2d}}.
\end{equation}
From the above analysis we can say that
\begin{equation}
 \lim_{N \to \infty}\Pr\left\{E^k_{\mathcal{I}^*}< 0\right\} = 0, \quad \forall\ k = k_0+1,\ldots, K. \label{eq:overfit_N_lim_EkIstar_<_0}
\end{equation}
Finally, using the union bound and the result in (\ref{eq:overfit_N_lim_EkIstar_<_0}) we can express the probability of $\mathcal{C}_1$ (\ref{eq:C1}) happening as
\begin{align}
    \Pr \left\{ \mathcal{C}_1\right \} =& \Pr \left \{ \bigcap_{k=k_0+1}^{K}E^k_{\mathcal{I}^*} > 0\right\} \nonumber\\
    &\geq 1-\sum_{k=k_0+1}^{K}\Pr \left\{ E^k_{\mathcal{I}^*} < 0\right\} \to 1 \label{eq:overfit_N_C1}
\end{align}
as $N \to \infty$.

$(b)$ \textit{Misfitting case} $(\mathcal{S} \not \subset \hat{\mathcal{S}}_{\text{EBIC}_\text{R}})$:
Let $\mathcal{I}_j \in \mathcal{I}^k_m$ be any misfitted subset of dimension $k$. Consider the following inequality
\begin{equation}
    \text{EBIC}_\text{R}(\mathcal{I}_j)>\text{EBIC}_\text{R}(\mathcal{S}), \quad \mathcal{I}_j \in \mathcal{I}^k_m. \label{eq:EFIC_S_less_EFIC_I_misfit_N_0}
\end{equation}
Denoting $\Delta = k - k_0$ and rearranging (\ref{eq:EFIC_S_less_EFIC_I_misfit_N_0}) we get
\begin{equation}
\begin{split}
    (N-k-2)\ln \left(\frac{\hat{\sigma}^2_{\mathcal{I}_j}}{\hat{\sigma}^2_{\mathcal{S}}}\right) + (1+2\zeta d)\Delta\ln N \\
    +\Delta \ln \left(\frac{\hat{\sigma}^2_0}{\hat{\sigma}^2_{\mathcal{S}}}\right)+ \Delta \ln \left(\frac{1}{2\pi}\right)  > 0.\label{eq:misfitting_N_inequality}
\end{split}
\end{equation}
Let $E^k_{\mathcal{I}_j}$ denote the entire left hand side of the inequality in (\ref{eq:misfitting_N_inequality}) and  $\mathcal{I}^*$ denote the subset that generates the minimum value of $E^k_{\mathcal{I}_j}$ among all such subsets of dimension $k$. Then we have
 \begin{equation}
     E^k_{\mathcal{I}^*} = \underset{\mathcal{I}_j \in \mathcal{I}^k_m}{\min} \left\{ E^k_{\mathcal{I}_j} \right\}, \quad  \quad j = 1,2,\ldots , T,
 \end{equation}
where  $T={{p}\choose{k}}$ if $k<k_0$ otherwise $T = {{p}\choose{k}}-{{p-k_0}\choose{\Delta}}$ if $k\geq k_0$.
The condition $\mathcal{C}_2$ in (\ref{eq:C2}) is satisfied as $N \to \infty$ under the event $     E^k_{\mathcal{I}^*} > 0, $ for all $ k=1,\ldots,K .$
Now, let $\mathbf{u}= \mathbb{E}[\mathbf{y}] = \mathbf{A}_\mathcal{S}\mathbf{x}_\mathcal{S}$. Using this, the ratio $\frac{\hat{\sigma}^2_{\mathcal{I}_j}}{\hat{\sigma}^2_{\mathcal{S}}}$ can be expanded as
\begin{align}
    \frac{\hat{\sigma}^2_{\mathcal{I}_j}}{\hat{\sigma}^2_{\mathcal{S}}} &= \frac{\mathbf{y}^T\boldsymbol{\Pi}^\perp_{\mathcal{I}_j}\mathbf{y} }{\mathbf{y}^T\boldsymbol{\Pi}^\perp_{\mathcal{S} }\mathbf{y}} = \frac{(\mathbf{u} + \mathbf{e})^T \boldsymbol{\Pi}^\perp_{\mathcal{I}_j}(\mathbf{u} + \mathbf{e}) }{\mathbf{e}^T\boldsymbol{\Pi}^\perp_{\mathcal{S} }\mathbf{e}}\nonumber\\
    & = \frac{\mathbf{u}^T \boldsymbol{\Pi}^\perp_{\mathcal{I}_j}\mathbf{u} + 2\sigma \sqrt{\mathbf{u}^T \boldsymbol{\Pi}^\perp_{\mathcal{I}_j}\mathbf{u}}\cdot  Z_j +\mathbf{e}^T\boldsymbol{\Pi}^\perp_{\mathcal{I}_j}\mathbf{e}}{\mathbf{e}^T\boldsymbol{\Pi}^\perp_{\mathcal{S} }\mathbf{e}},
    \label{eq:misfit_N_ratio_expanded}
\end{align}
where
\begin{equation}
    Z_j = \frac{\mathbf{u}^T \boldsymbol{\Pi}^\perp_{\mathcal{I}_j}\mathbf{e}}{\sigma \sqrt{\mathbf{u}^T \boldsymbol{\Pi}^\perp_{\mathcal{I}_j}\mathbf{u}}} \sim \mathcal{N}(0,1).
\end{equation}
Now
\begin{align}
    &\underset{1\leq j\leq T}{\min} \big \{\hat{\sigma}^2_{\mathcal{I}_j}/\hat{\sigma}^2_{\mathcal{S}}\big \}= \nonumber\\
    & \small \underset{1\leq j\leq T}{\min}\bigg \{ \mathbf{u}^T \boldsymbol{\Pi}^\perp_{\mathcal{I}_j}\mathbf{u}+ 2\sigma \sqrt{\mathbf{u}^T \boldsymbol{\Pi}^\perp_{\mathcal{I}_j}\mathbf{u}}\cdot  Z_j + \mathbf{e}^T\boldsymbol{\Pi}^\perp_{\mathcal{I}_j}\mathbf{e} \bigg \}\bigg /\mathbf{e}^T\boldsymbol{\Pi}^\perp_{\mathcal{S} }\mathbf{e} \nonumber\\
    & \geq \bigg [\underset{1\leq j\leq T}{\min}\big \{ \mathbf{u}^T \boldsymbol{\Pi}^\perp_{\mathcal{I}_j}\mathbf{u}\big\} + \sigma ^2\underset{1\leq j\leq T}{\min}\big\{\mathbf{e}^T\boldsymbol{\Pi}^\perp_{\mathcal{I}_j}\mathbf{e}/\sigma^2\big \} \nonumber\\
    &\quad -2\sigma  \sqrt{\underset{1\leq j\leq T}{\max}\big \{\mathbf{u}^T \boldsymbol{\Pi}^\perp_{\mathcal{I}_j}\mathbf{u}\big\}}\cdot   \underset{1\leq j\leq T}{\max}\big \{Z_j\big \}\bigg ]\bigg /\mathbf{e}^T\boldsymbol{\Pi}^\perp_{\mathcal{S} }\mathbf{e} .
\end{align}
In the misfitting scenario we have two cases: (i) $k<k_0$ (ii) $k\geq k_0$. We consider case (i) in our further analysis, which also encapsulates case (ii). For $k<k_0$ we have $\ln T = \ln \binom{p}{k} \approx kd\ln N$. Therefore, using the result in Lemma 2 we have the following lower bound under large-$N$ approximation
\begin{align}
    \underset{1\leq j\leq T}{\min}\bigg\{{\mathbf{e}^T\boldsymbol{\Pi}^\perp_{\mathcal{I}_j}\mathbf{e}/\sigma^2}\bigg\} =& \  {\mathbf{e}^T\mathbf{e}/\sigma^2-\underset{1\leq j\leq T}{\max}\bigg\{\mathbf{e}^T\boldsymbol{\Pi}_{\mathcal{I}_j}\mathbf{e}/\sigma^2\bigg\}}\nonumber\\
    &\geq \ N-2\psi'k d\ln N, \label{eq:lower-bound-chi2}
\end{align}
where $\psi' >1$ and $\mathbf{e}^T\mathbf{e}/\sigma^2 \approx N$ for large-$N$. Furthermore, from the result in Lemma \ref{lemma-Gaussian-bound}  we have the following upper bound
\begin{equation}
    \underset{1\leq j\leq T}{\max}\{Z_j\} \leq \sqrt{2\psi'k d\ln N}, \label{eq:upper-bound-Gaussian}
\end{equation}
where $\psi'>1$. Now, let $C_\text{min} = \underset{1\leq j\leq T}{\min}\big \{ \mathbf{u}^T \boldsymbol{\Pi}^\perp_{\mathcal{I}_j}\mathbf{u} \big \} $ and $C_\text{max} = \underset{1\leq j\leq T}{\max}\big \{ \mathbf{u}^T \boldsymbol{\Pi}^\perp_{\mathcal{I}_j}\mathbf{u} \big \} $. Also as $N\to \infty$ we can approximate $(N-k-2)\approx N$ and $\mathbf{e}^T\boldsymbol{\Pi}^\perp_{\mathcal{S} }\mathbf{e} \approx \sigma^2 N$. Using this, and the results in (\ref{eq:lower-bound-chi2}) and (\ref{eq:upper-bound-Gaussian}) we get
\begin{align}
     &\underset{1\leq j\leq T}{\min} \left \{N \ln  \left(\frac{\hat{\sigma}^2_{\mathcal{I}_j}}{\hat{\sigma}^2_{\mathcal{S}}}\right) \right\}  = N\ln \left[  \underset{1\leq j\leq T}{\min} \bigg \{\frac{\hat{\sigma}^2_{\mathcal{I}_j}}{\hat{\sigma}^2_{\mathcal{S}}}\bigg \} \right]\nonumber\\
     &\qquad \geq N\ln \bigg[\bigg \{C_\text{min} - 2\sigma \sqrt{C_\text{max}}\cdot  \sqrt{2\psi'kd\ln N} \nonumber\\
     & \hspace{2.5cm} +\sigma^2 \left(N-2\psi'k d\ln N\right)\bigg\}\bigg /\sigma^2 N\bigg]\label{eq:min-bound-log-ratio-missfit-N}.
\end{align}
Now, observe that $C_\text{min} = \mathbf{u}^T \boldsymbol{\Pi}^\perp_{\mathcal{I}^{*}}\mathbf{u} = \mathbf{x}_\mathcal{S}^T\mathbf{A}^T_\mathcal{S} \boldsymbol{\Pi}^\perp_{\mathcal{I}^{*}}\mathbf{A}_\mathcal{S}\mathbf{x}_{\mathcal{S}}$. Since, we are in the misfitting scenario, from Lemma \ref{lemma-identifiability}, in Appendix \ref{appendix_lemmas-2-to-4}, we can express $C_{\min} = Nb_{\min} $ where $b_{\min} =\mathcal{O}(1)>0$. Similarly, $C_{\max} = Nb_{\max}$ where $b_{\max}=\mathcal{O}(1)>0$ and $0< b_{\min} \leq b_{\max}$. Hence, we can rewrite (\ref{eq:min-bound-log-ratio-missfit-N}) as
\begin{align}
    &\underset{1\leq j\leq T}{\min} \left \{N \ln  \left(\frac{\hat{\sigma}^2_{\mathcal{I}_j}}{\hat{\sigma}^2_{\mathcal{S}}}\right) \right\} \geq \nonumber\\
    & N\ln \left(1+\frac{b_\text{min}}{\sigma^2}-\frac{2\sqrt{b_\text{max}}}{\sigma}\sqrt{\frac{2\psi'kd\ln N}{N}} -\frac{2\psi'k d\ln N}{N} \right)\nonumber\\
    & \approx N \ln \left(1+\frac{b_\text{min}}{\sigma^2} \right)
\end{align}
as $N$ grows large. For $k<k_0$, we get $\Delta <0$, therefore, in this case we have
\begin{equation}
    \begin{split}
        E^k_{\mathcal{I}^*} \geq  N \ln \left(1+ \frac{b_\text{min}}{\sigma^2}\right) - |\Delta|(1+2\zeta d)\ln N\\ 
        -|\Delta| \ln \left(\frac{\hat{\sigma}^2_0}{2\pi\hat{\sigma}^2_{\mathcal{S}}}\right) \to \infty \label{eq:Ek-underfitting-to-infty}
    \end{split}
\end{equation}
as $N \to \infty$  for all $k=1, \ldots, K$, since $N\ln (1+b_\text{min}/\sigma^2)$ is the dominating term as it tends to infinity much faster than the $\ln N$ term and $\lim_{N \to \infty} \hat{\sigma}^2_{0} = \text{const.} >0$ (see Appendix \ref{appendix_sigma2_0_analysis}) and $\lim_{N \to \infty} \hat{\sigma}^2_\mathcal{S} = \sigma^2$ (see Appendix \ref{appendix_sigma2_I_analysis}). From the above analysis we can say that
\begin{equation}
 \lim_{N \to \infty}\Pr\left\{E^k_{\mathcal{I}^*}< 0\right\} = 0, \quad \forall\ k = 1, \ldots, K.  \label{eq:misfit_N_EkIstar_0}
\end{equation}
Finally, using the union bound and the result in (\ref{eq:misfit_N_EkIstar_0}) we can express the probability of $\mathcal{C}_2$ (\ref{eq:C2}) happening as
\begin{align}
    \Pr \left\{ \mathcal{C}_2\right \} &= \Pr \left \{ \bigcap_{k=1}^{K}E^k_{\mathcal{I}^*} > 0\right\} \nonumber\\
    &\geq 1-\sum_{k=1}^{K}\Pr \left\{ E^k_{\mathcal{I}^*} < 0\right\} \to 1 \quad \text{as} \quad N \to \infty. \label{eq:misfit_N_C2}
\end{align}
From (\ref{eq:overfit_N_C1}) and (\ref{eq:misfit_N_C2}) we can conclude that EBIC$_\text{R}$ is consistent as $N\to \infty$, which proves Theorem 2.

\subsection{Discussion on the Hyperparameter $\zeta$}
If $\zeta$ is too large, it will lead to underfitting issues. This is evident from (\ref{eq:Ek-underfitting-to-infty}) where a large value of $\zeta$ may force the overall sum to become negative especially for smaller $N$ values. On the contrary, if $\zeta$ is too small, it will lead to overfitting issues as the penalty may not be sufficiently large to compensate for the overparameterization due to large parameter space.

\section{Predictor selection algorithms}
In the high-dimensional scenario, when $p$ is large, it is infeasible to perform model selection in the conventional manner. For a design matrix with parameter dimension $p$, the number of possible candidate models is $2^p-1$. Hence, the candidate model space grows exponentially with $p$ and we cannot afford to calculate model score for all possible models. Therefore, to perform model selection, we combine a model selection criterion with a predictor selection (support recovery) algorithm such as OMP or LASSO (least absolute shrinkage and selection operator) \cite{lasso_tibshirani1996}.
The goal of predictor selection is to pick a subset of important predictors from the entire set of $p$ predictors. 
In this context, the most important predictors refer to the positions of the nonzero elements of the input signal $\mathbf{x}$.
Thus, predictor selection reduces the cardinality of the candidate model space to some upper bound $K$ such that $k_0\leq K\ll N$ under the assumption of a sparse parameter vector.
This enables us to apply the model selection criterion on the smaller set of candidate models to pick the best model.
The OMP algorithm is shown in Algorithm \ref{alg:OMP}. To perform model selection, we combine OMP with EBIC$_\text{R}$ as shown in Algorithm \ref{alg:OMP_ebicR}. 
\begin{algorithm}[t]
\caption{OMP with $K$ iterations}\label{alg:OMP}
\begin{algorithmic}
\State \textbf{Inputs:} Design matrix $\mathbf{A}$, measurement vector $\mathbf{y}$.
\State \textbf{Initialization:} $\lVert \mathbf{a}_j\rVert_2 = 1\ \forall j$, $\mathbf{r}^0=\mathbf{y}$, $\mathcal{S}^0_\text{OMP} = \emptyset$
\For{$i=1 \text{\ to\ } K$}  
	\State Find next column index: $d^i = \underset{j}{\arg\max} \big|\mathbf{a}_j^T\mathbf{r}^{i-1}\big|$
	\State Add current index: $\mathcal{S}^i_{\textrm{OMP}} = \mathcal{S}^{i-1}_{\textrm{OMP}} \cup \{d^i\}$
	\State Update residual: $\mathbf{r}^i = \left(\mathbf{I}_N-\mathbf{\Pi}_{\mathcal{S}^i_{\textrm{OMP}}}\right)\mathbf{y}$
\EndFor
\State \textbf{Output:} OMP generated index sequence $\mathcal{S}^K_\text{OMP}$
\end{algorithmic}
\end{algorithm}
\begin{algorithm}[b]
\caption{Model selection combining EBIC$_\text{R}$ with OMP}\label{alg:OMP_ebicR}
\begin{algorithmic}
\State Run OMP for $K$ iterations to obtain  $\mathcal{S}^K_\text{OMP}$
\For{$k=1 \text{\ to\ } K$}  
\State $\mathcal{I} = \mathcal{S}^k_\text{OMP}$
\State Compute EBIC$_\text{R}(\mathcal{I})$
\EndFor
\State Estimated true support: $\hat{\mathcal{S}}_{\text{EBIC}_\text{R}} = \underset{\mathcal{I}}{\arg\min} \{$EBIC$_\text{R}(\mathcal{I})\}$
\end{algorithmic}
\end{algorithm}
LASSO is a shrinkage method for variable selection/estimation in linear regression models developed by Tibshirani \cite{lasso_tibshirani1996}. Given the linear model in (\ref{eq:LR_model_1}), the LASSO solution for $\mathbf{x}$ for a particular choice of the regularization parameter $\lambda \geq 0$ is obtained as
\begin{equation}
    \hat{\mathbf{x}}_\text{lasso}(\lambda) = \underset{\mathbf{x}\in \mathbb{R}^p}{\min} \left\{ \frac{1}{2N} \lVert \mathbf{y} - \mathbf{A}\mathbf{x} \rVert^2_2 + \lambda \lVert \mathbf{x}\rVert_1  \right\}\label{eq:lasso_solution},
\end{equation}
where $\lVert \cdot\rVert_1$ denotes the $l_1$ norm. The parameter $\lambda$ determines the level of sparsity. When $\lambda\to \infty$ the objective function in (\ref{eq:lasso_solution}) attains the minimum with $\hat{\mathbf{x}}_\text{lasso}(\lambda)$ being a zero vector. As we gradually lower the $\lambda$ value, the number of non-zero components in $\hat{\mathbf{x}}_\text{lasso}(\lambda)$ starts increasing. Model selection combining LASSO and EBIC$_\text{R}$ can be performed as shown in Algorithm \ref{alg:lasso_ebic_R}. Gradually decrease $\lambda$ from a high value so that the number of non-zero components in $\hat{\mathbf{x}}_\text{lasso}(\lambda)$ gradually increases. Therefore, for each decreasing unique value of $\lambda$ say $\lambda_i$, we acquire a different solution $\hat{\mathbf{x}}_\text{lasso}(\lambda_i)$, with increasing support and thus obtaining a sequence of candidate models with maximum cardinality $K$. The value of EBIC$_\text{R}$ is computed for each of the candidate models and the model corresponding to the smallest EBIC$_\text{R}$ score is selected as the final model. A most useful method for solving LASSO in our context is the (modified) least angle regression (LARS) algorithm \cite{LARS_Efron2004}, since it also provides the required sequence of regularization parameters for which the support changes. 


\begin{algorithm}[t]
\caption{Model selection combining EBIC$_\text{R}$ with LASSO}\label{alg:lasso_ebic_R}
\begin{algorithmic}
\State Compute LASSO estimates $\{\hat{\mathbf{x}}_\text{lasso}(\lambda_1),\ldots,\hat{\mathbf{x}}_\text{lasso}(\lambda_{K_\text{max}})\}$ where $card(supp\left(\hat{\mathbf{x}}_\text{lasso}(\lambda_{K_\text{max}})\right)) = K$
\For{$i = 1 \text{\ to \ } K_{\max}$}
\State $\mathcal{I} = supp\left(\hat{\mathbf{x}}_\text{lasso}(\lambda_i)\right)$
\State Compute EBIC$_\text{R}(\mathcal{I})$
\EndFor
\State Estimated true support: $\hat{S}_{\text{EBIC}_\text{R}} = \underset{\mathcal{I}}{\arg\min} \ \{$EBIC$_\text{R}(\mathcal{I})\}$
\end{algorithmic}
\end{algorithm}

\section{Simulation Results}
In this section, we provide numerical simulation results to illustrate the empirical performance of EBIC$_\text{R}$. The performance of EBIC$_\text{R}$ is compared with the `oracle', EBIC, EFIC and MBT.
However, the performance comparison with the RRT \cite{RRT-2018} method is dropped since it behaves quite similar to MBT (see \cite{gohain2020_MBT} for details).
The `oracle' criterion assumes \textit{a priori} knowledge of the true cardinality $k_0$. Thus, the model selection performance of the `oracle' provides the upper bound on the model selection performance that can be achieved using a particular predictor selection algorithm and for a given set of data settings. Additionally, we also provide simulation results to highlight the drawbacks of classical  methods for model selection in high-dimensional linear regression models with a sparse parameter vector.
\subsection{General Simulation Setup}
In the simulations, we consider the model $\mathbf{y=Ax +e}$, where the design matrix $\mathbf{A}\in \mathbb{R}^{N\times p}$ is generated with independent entries following normal distribution $\mathcal{N}(0,1)$. Since $\mathbf{x}$ is assumed to be sparse, we choose $k_0 =5$. Furthermore, without loss of generality, we assume that the true support is $\mathcal{S} = [1,2,3,4,5]$, therefore, $\mathbf{x}_\mathcal{S} = [x_1,x_2,x_3,x_4,x_5]^T$ and $\mathbf{A}_\mathcal{S} = [\mathbf{a}_1,\mathbf{a}_2,\mathbf{a}_3,\mathbf{a}_4,\mathbf{a}_5]$. This implies that the elements of $\mathbf{x}$ follows $x_k\neq 0$ for $k=1,\ldots,k_0$ and $x_k =0$ for $k>k_0$. 
The SNR in dB is SNR (dB) = $10\log_{10}(\sigma_s^2/\sigma^2)$, where $\sigma_s^2$ and $\sigma^2$ denote signal and true noise power, respectively. The signal power is computed as $\sigma_s^2 = ||\mathbf{A}_\mathcal{S}\mathbf{x}_\mathcal{S}||^2_2/N$. Based on $\sigma_s^2$ and the chosen SNR (dB), the noise power is set as $\sigma^2 = \sigma_s^2/10^{\textrm{SNR (dB)}/10} $. Using this $\sigma^2$, the noise vector $\mathbf{e}$ is generated following $\mathcal{N}(\mathbf{0},\sigma^2\mathbf{I}_{N})$. The probability of correct model selection (PCMS) is estimated over $1000$ Monte Carlo trials. To maintain randomness in the data, a new design matrix $\mathbf{A}$ is generated at each Monte Carlo trial. OMP is used for predictor selection for its simplicity and wider range of applicability.
\subsection{Tuning Parameter Selection}
\begin{figure}[t]
    \centering
    \includegraphics[trim={2cm 0 2cm 0.5cm},clip,scale=0.25]{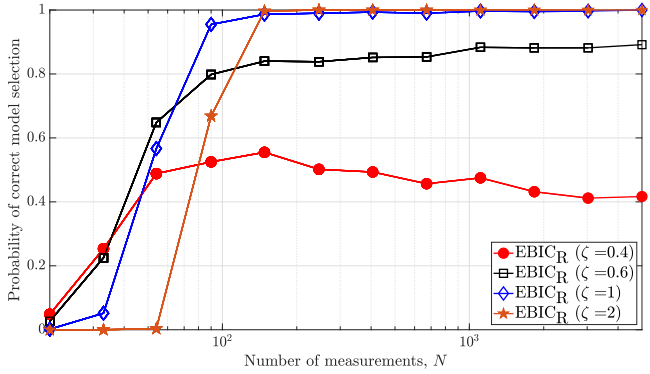}    \caption{PCMS vs $N$ with $\mathbf{x}_\mathcal{S} = [1,1,1,1,1]$, SNR = 5 dB, $p=N^d$ and $d=1.1$.}
    \label{fig:tuning_parameter_comparison}
\end{figure}
An important step in model selection is the choice of the tuning parameter. As mentioned earlier, too small or large values of the tuning parameter can cause severe performance degradation in certain scenarios. Fig. \ref{fig:tuning_parameter_comparison} shows a performance comparison of EBIC$_\text{R}$ for four different values of $\zeta$ (0.4, 0.6, 1, and 2). Here, we set $p = N^d$ where $d=1.1$. Hence, from Theorem \ref{theorem-large-N} we require $\zeta>1-1/2d=0.55$ to achieve consistency. From the figure we see that for $\zeta = 0.4$, the performance of EBIC$_\text{R}$ degrades after a certain point with increasing $N$, which justifies the theory. For all other $\zeta>0.55$, the performances improve with increasing $N$. For $\zeta=0.6$, which is very close to the lower bound, the convergence to correct selection probability one is slow and will require a very large sample size. For, $\zeta = 2$, the performance suffers (due to underfitting) in the low $N$ regime, but do achieve perfect selection as $N$ increases. In this case, $\zeta =1$ provides a much better overall performance for a broader range of $N$. A similar trend as in EBIC$_\text{R}$ is observed even in EBIC and EFIC for different choices of $\gamma$ and $c$. Hence, to maintain fairness, the following tuning parameter settings are considered for further analysis: $\zeta = 1$ (EBIC$_\text{R}$), $c = 1$ (EFIC) and  $\gamma = 1$ (EBIC). For MBT \cite{gohain2020_MBT}, $\lim\limits_{N\to \infty}\text{PCMS} \to 1$ as $\beta \to 1$. Hence, we choose $\beta = 0.999$.

\subsection{Model Selection with Classical Methods in High-Dimensional Setting}\label{sec:MS_with_classical_methods}
\begin{figure}[b]
    \centering
    \includegraphics[trim={2cm 0 2cm 1cm},clip,scale=0.25]{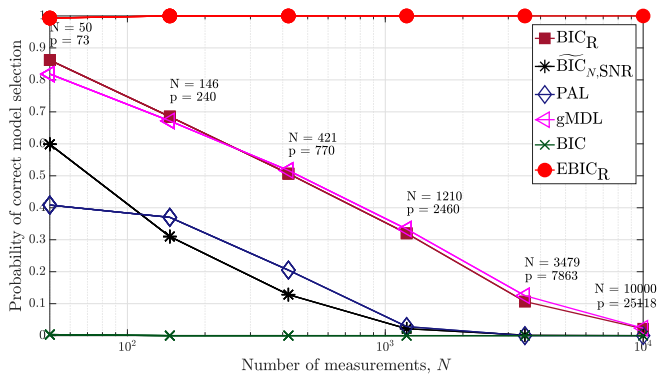}    
    \caption{The PCMS versus $N$ for SNR = 30 dB with $\mathbf{x}_\mathcal{S} = [5,4,3,2,1]$ and $p=N^{d}$ where $d=1.1$. }
    \label{fig:P_vs_N_classical}
\end{figure}

In this section, we present simulations results for model selection using classical methods in high-dimensional linear regression models and compare their performances with EBIC$_\text{R}$. The purpose of these results is to highlight the limitations of the classical methods in dealing with large-$p$ small-$N$ scenarios. The classical methods used here are BIC \cite{BIC_Schwarz1978}, $\widetilde{\text{BIC}}_{N,\text{SNR}}$\cite{Stoica_BIC_SNR}, BIC$_\text{R}$\cite{GOHAIN-BIC-R}, gMDL \cite{gMDL_Hansen2001}, and PAL \cite{PAL_Stoica2013}. 

In the simulation, we consider the true parameter vector to be $\mathbf{x}_\mathcal{S} = [5,4,3,2,1]^T$. Fig. \ref{fig:P_vs_N_classical} presents the plot for PCMS versus $N$ for SNR = 30 dB with $p=N^d$ where $d=1.1$.
The figure shows that EBIC$_\text{R}$ ($\zeta=1$) clearly surpasses the classical methods with huge differences in performance. 
In general, when $p$ is fixed and $N\to \infty$, the classical methods are consistent \cite{GOHAIN-BIC-R}. However, when $p$ is varying and grows exponentially with $N$, the consistency attribute does not hold any longer, hence, we see the decreasing performance trend in Fig. \ref{fig:P_vs_N_classical}. 
\begin{figure}[b]
    \centering
    \includegraphics[trim={2cm 0 2cm 1cm},clip,scale=0.25]{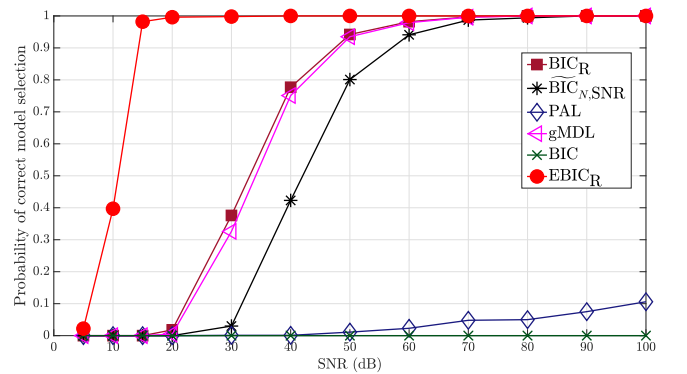}
    \caption{The PCMS versus SNR (dB) for $N = 100$, $p = 500$ and  $\mathbf{x}_\mathcal{S} = [5,4,3,2,1]$.}
    \label{fig:P_vs_SNR_classical}
\end{figure}
\begin{figure*}[h]
    \centering
    \begin{subfigure}[h]{0.5\textwidth}
        \centering
        \includegraphics[trim={2cm 0 2cm 0.5cm},clip,scale=0.23]{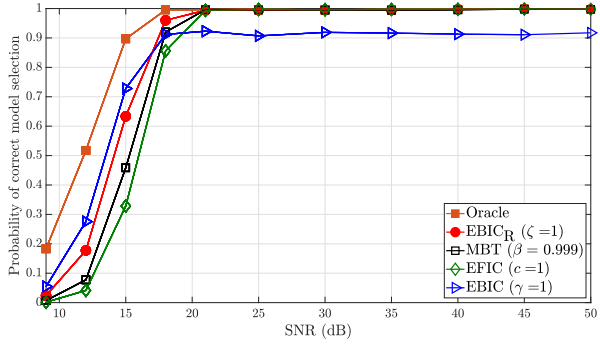}
        \caption{$\mathbf{x}_\mathcal{S} = [0.05, 0.04, 0.03, 0.02, 0.01]^T$}
        \label{fig:P_vs_SNR_point01}
    \end{subfigure}%
    ~ 
    \begin{subfigure}[h]{0.5\textwidth}
        \centering
        \includegraphics[trim={2cm 0 2cm 0.5cm},clip,scale=0.23]{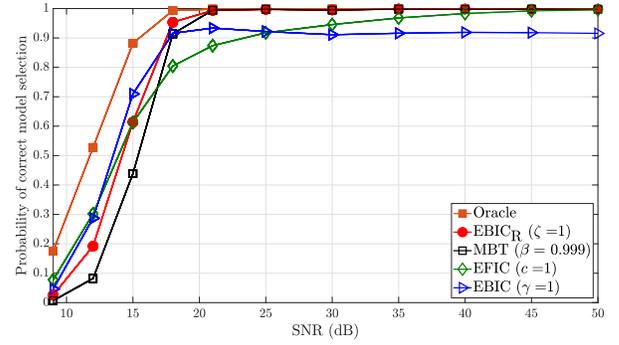}
        \caption{$\mathbf{x}_\mathcal{S} = [50, 40, 30, 20, 10]^T$}
        \label{fig:P_vs_SNR10}
    \end{subfigure}
    \caption{The PCMS versus SNR (dB) for $N = 55$ and $p = 1000$.}
    \label{fig:P_vs_SNR}
\end{figure*}

Fig. \ref{fig:P_vs_SNR_classical} illustrates the PCMS versus SNR in dB for fixed $N = 100$ and $p=500$. This gives $d = \log(p)/\log(N) \approx 1.35$, hence, $\zeta >1-1/2d\approx 0.63$.
The first major observation from the figure is that EBIC$_\text{R}$ ($\zeta=1$) clearly outperforms all the classical methods by a huge margin. Secondly, for the considered setting, the performances of BIC$_\text{R}$ and gMDL are quite similar followed by  $\widetilde{\text{BIC}}_{N,\text{SNR}}$. The criteria BIC$_\text{R}$, gMDL and $\widetilde{\text{BIC}}_{N,\text{SNR}}$ do achieve convergence to detection probability one but at the expense of very high values of SNR. The performances of PAL and BIC are extremely poor in this case, even in the high-SNR regions.

\subsection{Model Selection with the Latest Methods in High-Dimensional Setting}
In Section \ref{sec:MS_with_classical_methods}, we highlighted the drawbacks of classical methods in model selection under the high-dimensional setting. We observed that the performance of the classical methods collapses when $p$ grows exponentially with $N$ and the consistency property breaks down. In this section, we present simulation results for model selection comparing EBIC$_\text{R}$ to the existing state-of-the-art methods, designed to deal with the large-$p$ small-$N$  scenarios. 

\subsubsection{Model Selection versus SNR}
To highlight the scale-invariant and consistent behaviour of EBIC$_\text{R}$, we consider two scenarios. In the first scenario, we assume the true parameter vector to be $\mathbf{x}_\mathcal{S} = [0.05,0.04,0.03,0.02,0.01]^T$ 
and in the second scenario, we assume $\mathbf{x}_\mathcal{S} = [50,40,30,20,10]^T$.
Note that in the simulations we compute the noise variance $\sigma^2$ based on the chosen SNR level and the current signal power value $\sigma^2_s = \lVert \mathbf{A}_\mathcal{S}\mathbf{x}_\mathcal{S} \rVert^2_2\big /N$. To simulate the probability of correct model selection versus SNR in a high-dimensional setting we fixed $N=55$ and $p=1000$. This gives $d = \log(p)/\log(N) \approx 1.724$, hence, $\zeta >1-1/2d \approx 0.71$.
\begin{figure}[b]
    \centering
    \includegraphics[trim={2cm 0 2cm 0.5cm},clip,scale=0.25]{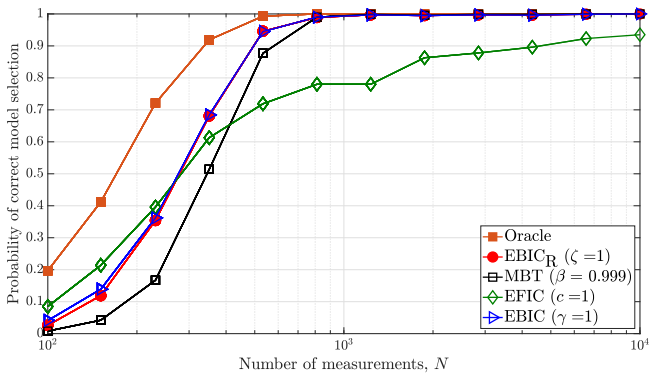}
    \caption{The PCMS versus $N$ for SNR = $6$ dB, $p = 1000$ and $\mathbf{x}_\mathcal{S}=[50,40,30,20,10]^T$.}
    \label{fig:P_vs_N}
\end{figure}

Fig. \ref{fig:P_vs_SNR} shows the empirical PCMS versus SNR (dB). Fig. \ref{fig:P_vs_SNR_point01} and Fig. \ref{fig:P_vs_SNR10} correspond to $\mathbf{x}_\mathcal{S} = [0.05,0.04,0.03,0.02,0.01]$ and  $\mathbf{x}_\mathcal{S} = [50,40,30,20,10]$, respectively. Both the figures depict fixed $N$ increasing SNR scenario. Comparing the figures, the first clear observation is that unlike the other criteria, the behaviour of EFIC is not identical for the two different $\mathbf{x}_\mathcal{S}$ given that the other parameters viz, $N$, $p$ and $k_0$ are constant and the performance is evaluated for the same SNR range. This illustrates the scaling problem present in EFIC that leads to either high underfitting or overfitting issues. This behavior or EFIC can be explained as follows. 
The data dependent penalty term (DDPT) of EFIC is $\text{DDPT}=-(k+2)\ln \lVert \mathbf{\Pi}^\perp_\mathcal{I}\mathbf{y}\rVert^2_2$, whose overall value depends on the value $\lVert \mathbf{\Pi}^\perp_\mathcal{I}\mathbf{y}\rVert^2_2$, which in turn is influenced by the signal and noise powers $\sigma^2_s$ and $\sigma^2$, respectively. If $\lVert\mathbf{\Pi}^\perp_\mathcal{I}\mathbf{y}\rVert_2^2 \ll 1$, then $\text{DDPT}\gg 0$, which may blow the overall penalty to a large value leading to underfitting issues. This is most likely the case when $\mathbf{x}_\mathcal{S} = [0.05,0.04,0.03,0.02,0.01]^T$ (Fig. \ref{fig:P_vs_SNR_point01}). On the contrary if $\lVert\mathbf{\Pi}^\perp_\mathcal{I}\mathbf{y}\rVert^2_2 \gg 1$, then $\text{DDPT}\ll 0$, thus lowering the overall penalty leading to overfitting issues (when $\mathbf{x}_\mathcal{S} = [50,40,30,20,10]^T$, Fig. \ref{fig:P_vs_SNR10}).
The second major observation is that EBIC is inconsistent when SNR is high but $N$ is small and fixed. This behaviour of EBIC is already  reported in \cite{EFIC2018}. In general, EFIC, MBT (for $\beta \to 1$) and EBIC$_\text{R}$ are consistent for increasing SNR scenarios given that $N$ is fixed, but while EBIC$_\text{R}$ and MBT are invariant to data scaling EFIC is not.
\begin{figure}[b]
    \centering
    \includegraphics[trim={2cm 0 2cm 0.5cm},clip,scale=0.25]{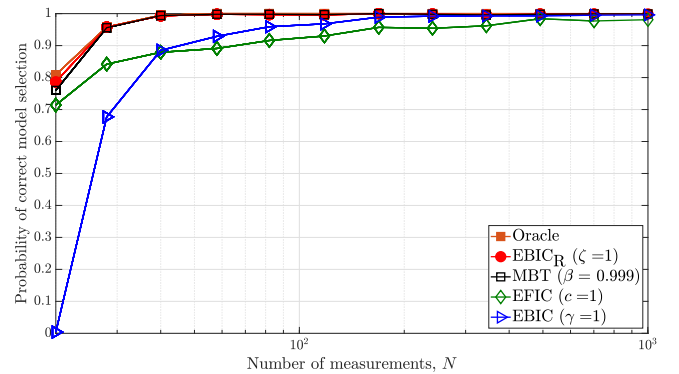}
    \caption{The PCMS versus $N$ for SNR = $25$ dB, $p = N^d$, $d=1.3$ and $\mathbf{x}_\mathcal{S}=[50,40,30,20,10]^T$.}
    \label{fig:P_vs_N_high_SNR}
\end{figure}
\subsubsection{Model Selection versus $N$}

Fig. \ref{fig:P_vs_N} illustrates the empirical PCMS versus $N$ for SNR = 6 dB, $p=1000$ and $\mathbf{x}_\mathcal{S}=[50,40,30,20,10]^T$. It depicts a low-SNR increasing $N$ scenario. 
It is clearly seen that compared to the other criteria, EFIC suffers from the scaling issue and requires a large sample size to achieve detection probability one. 
Among all the criteria, the performance of EBIC and EBIC$_\text{R}$ are closest to the oracle.
Furthermore, observe that the performance of EBIC$_\text{R}$ and EBIC are more or less alike for the current setting. This is primarily because the SNR is low (6dB) hence the $(k+2)\ln(\hat{\sigma}^2_0/\hat{\sigma}^2_\mathcal{I})$ term of EBIC$_\text{R}$ behaves very close to a $\mathcal{O}(1)$ quantity for $k\geq k_0$. Thus, for low SNR scenarios, the penalties of EBIC and EBIC$_\text{R}$ are similar and as such the behaviour of these two criteria overlaps in this case. However, note that this is not true in the high-SNR cases, which will be evident from the discussion following Fig. \ref{fig:P_vs_N_high_SNR}.

The plots shown in Fig. \ref{fig:P_vs_SNR} and Fig. \ref{fig:P_vs_N} represent fixed-$N$ increasing-SNR and low-SNR increasing-$N$ scenarios, respectively. In Fig. \ref{fig:P_vs_N_high_SNR}, we present a high-SNR increasing-$N$ case. Here, we consider a varying parameter space such that $p=N^d$ where $d=1.3$.
It is clearly observed that for high-SNR scenarios, EBIC$_\text{R}$ and MBT provide much faster convergence to oracle behaviour as compared to EBIC that requires higher sample size to achieve detection probability one. Furthermore, we also notice that EFIC suffers from a higher false selection error and performs worse than EBIC in a certain region of the sample size. This clearly shows the effects of scaling in the behaviour of EFIC.


\section{Conclusion}
In this paper, we provided a new criterion, which is an extension of BIC$_\text{R}$, to handle model selection in sparse high-dimensional linear regression models employing sparse methods for predictor selection. The extended version is named as EBIC$_\text{R}$, where the subscript `R' stands for robust and it is a scale-invariant and consistent model selection criterion. Additionally, we analytically examined the behaviour of EBIC$_\text{R}$ as $\sigma^2 \to 0$ and as $N\to \infty$. In both cases, it is shown that the probability of detecting the true model approaches one. The paper further highlighted the data scaling issue present in EFIC, which is a consistent criterion for both large sample size and high-SNR scenarios. Extensive simulation results show that the performance of EBIC$_\text{R}$ is either similar or superior to that of EBIC, EFIC and MBT.


\appendices
\section{}
\label{appendix_lemma_1}

\begin{lemma}\label{lemma-1}Let $\mathbf{y}$ be a $N\times 1$ dimensional vector following $\mathbf{y} \sim \mathcal{N}(\boldsymbol{\mu},\sigma^2\mathbf{I}_N)$ and $\mathbf{\Pi}$ be a $N \times N$ symmetric, idempotent matrix with rank$(\mathbf{\Pi}) = r$. Then the ratio $\mathbf{y}^T\mathbf{\Pi} \mathbf{y}/\sigma^2$ has a non-central chi-square distribution $\chi^2_r(\lambda)$ with $r$ degrees of freedom and non-centrality parameter $\lambda = \boldsymbol{\mu}^T \mathbf{\Pi}\boldsymbol{\mu}/\sigma^2$ (see, e.g.,  Chapter 5 of \cite{mathai1992quadratic}).
\end{lemma}

\section{statistical analysis of the factor $\hat{\sigma}^2_0$}\label{appendix_sigma2_0_analysis}
From the generating model (\ref{eq:LR_model_1}), the true data vector follows $\mathbf{y} \sim \mathcal{N}\left(\mathbf{A}_\mathcal{S}\mathbf{x}_\mathcal{S},\sigma^2\mathbf{I}_N\right)$. Consider the factor $\hat{\sigma}^2_0$, which is defined as
\begin{equation}
    \hat{\sigma}^2_0 = \frac{\lVert \mathbf{y}\rVert^2_2}{N} =\left(\frac{\sigma^2}{N}\right) \frac{ \mathbf{y}^T\mathbf{I}_N \mathbf{y}}{\sigma^2}.
\end{equation}
From Lemma \ref{lemma-1} in Appendix \ref{appendix_lemma_1} we have 
\begin{equation}
    \frac{ \mathbf{y}^T\mathbf{I}_N \mathbf{y}}{\sigma^2} \sim \chi^2_N(\lambda) \ \text{where} \ \lambda = \frac{\lVert \mathbf{A}_\mathcal{S}\mathbf{x}_\mathcal{S} \rVert^2_2}{\sigma^2}.
\end{equation}
This implies that $\left(\frac{N}{\sigma^2}\right) \hat{\sigma}^2_0 \sim  \chi^2_N(\lambda).$
Therefore, the mean and variance of $\hat{\sigma}^2_0$ are:
\begin{equation}
\begin{split}
    &\mathbb{E}[\hat{\sigma}^2_0] = \frac{\sigma^2}{N}(N +\lambda) = \sigma^2 + \frac{\lVert \mathbf{A}_\mathcal{S}\mathbf{x}_\mathcal{S} \rVert^2_2}{N} \\
    & \text{Var}[\hat{\sigma}^2_0] = 2\frac{\sigma^4}{N^2}(N+2\lambda) = 2\frac{\sigma^4}{N} + 4 \frac{\sigma^2}{N^2} \lVert \mathbf{A}_\mathcal{S}\mathbf{x}_\mathcal{S} \rVert^2_2.
\end{split}
\end{equation}
Hence, for a fixed $N$,
\begin{align}
    \lim_{\sigma^2 \to 0} \mathbb{E}[\hat{\sigma}^2_0] = \frac{\lVert \mathbf{A}_\mathcal{S}\mathbf{x}_\mathcal{S} \rVert^2_2}{N} \quad \& \quad  
    \lim_{\sigma^2 \to 0} \text{Var}[\hat{\sigma}^2_0] = 0.
\end{align}
Further, when SNR or $\sigma^2$ is fixed, using the assumption $\lim_{N\to \infty} \left\{ \frac{\mathbf{A}_\mathcal{S}^T\mathbf{A}_\mathcal{S}}{N}\right\} = \mathbf{M}_\mathcal{S}$ we get
\begin{align}
        \lim_{N \to \infty} \mathbb{E}[\hat{\sigma}^2_0] = \sigma^2 + \mathbf{x}_\mathcal{S}^T\mathbf{M}_\mathcal{S}\mathbf{x}_\mathcal{S} \quad \& \quad  
    \lim_{N \to \infty} \text{Var}[\hat{\sigma}^2_0] = 0,
\end{align}
where $\mathbf{M}_\mathcal{S}$ is a bounded positive definite matrix and as such $\mathbf{x}_\mathcal{S}^T\mathbf{M}_\mathcal{S}\mathbf{x}_\mathcal{S} = \mathcal{O}(1)$ as $N$ grows large.

\section{statistical analysis of  $\hat{\sigma}^2_\mathcal{I}$ when $\mathcal{S}\subseteq \mathcal{I}$}\label{appendix_sigma2_I_analysis}
The noise variance estimate under hypothesis $\mathcal{H}_\mathcal{I}$ can be rewritten as
\begin{align}
    \hat{\sigma}^2_\mathcal{I}  = \left(\frac{\sigma^2}{N}\right) \frac{\mathbf{y}^T \boldsymbol{\Pi}^\perp_\mathcal{I}\mathbf{y}}{\sigma^2}.
\end{align}
The true model $\mathbf{u} = \mathbf{A}_\mathcal{S}\mathbf{x}_\mathcal{S}$ lies in a linear subspace spanned by the columns of $\mathbf{A}_\mathcal{S}$. Consequently, for $\mathcal{I} \supseteq \mathcal{S}$ we have $\boldsymbol{\Pi}^\perp_\mathcal{I}\mathbf{u} = \mathbf{0}$. This implies that $\mathbf{y}^T \boldsymbol{\Pi}^\perp_\mathcal{I} \mathbf{y} = \mathbf{e}^T\boldsymbol{\Pi}^\perp_\mathcal{I} \mathbf{e}$.
Thus we have,
\begin{align}
    \frac{\mathbf{y}^T \boldsymbol{\Pi}^\perp_\mathcal{I}\mathbf{y}}{\sigma^2} = \frac{\mathbf{e}^T \boldsymbol{\Pi}^\perp_\mathcal{I}\mathbf{e}}{\sigma^2} \sim \chi^2_{N-k}\ \text{(Using Lemma \ref{lemma-1})},\label{eq:distr_greater_k0}
\end{align}
where $k = card(\mathcal{I})\geq k_0$. This implies that $     \left(\frac{N}{\sigma^2}\right)\hat{\sigma}^2_\mathcal{I} \sim \chi^2_{N-k}.$
Therefore, the mean and variance of $\hat{\sigma}^2_\mathcal{I}$ for $\mathcal{I} \supseteq \mathcal{S}$ are:
\begin{equation}
    \mathbb{E}[\hat{\sigma}^2_\mathcal{I}] = \frac{\sigma^2}{N}(N-k) \quad \& \quad \text{Var}[\hat{\sigma}^2_\mathcal{I}] = 2\frac{\sigma^4}{N^2}(N-k).
\end{equation}
Hence, when $\sigma^2$ is a constant,
\begin{equation}
    \lim_{N \to \infty}\mathbb{E}[\hat{\sigma}^2_\mathcal{I}] = \sigma^2 \quad \& \quad \lim_{N \to \infty}\text{Var}[\hat{\sigma}^2_\mathcal{I}] = 0.
\end{equation}

\section{}
\label{appendix_lemmas-2-to-4}
\begin{lemma}\label{lemma-Chi2-bound}
Let $Z_\text{max} = \underset{i}{\max}\big \{Z_i\big \}_{i=1}^{m}$ where $Z_1,Z_2,\ldots,Z_m$ is a sequence of identically distributed random variables (not necessarily independent) having a Chi-square distribution with $k$ degrees of freedom where $k<m$. Then $Z_\text{max} \leq  k + 2\sqrt{k\psi \ln m} + 2\psi \ln m$ for some constant $\psi >1$ with probability approaching one as $m \to \infty $. 
\end{lemma}

\textit{Proof:}
From the union bound we have
\begin{equation}
    \Pr \left(Z_\text{max} \leq \eta\right) \geq 1-m\Pr \left(Z_i \geq \eta\right).\label{eq:max-union-bound-overfit-N}
\end{equation}
Since $Z_i \sim \chi^2_k$, then from the Chi-square tail bound (Lemma 1 of \cite{laurent2000adaptive}) we have the following result
\begin{equation}
    \Pr \left(Z_i \geq k +2\sqrt{k t} + 2t\right) \leq e^{-t}.\label{eq:chi-square-bound-origin}
\end{equation}
Setting $t = \psi \ln m$ in (\ref{eq:chi-square-bound-origin}) where $\psi >1$ we get 
\begin{equation}
    \Pr \left(Z_i \geq k + 2\sqrt{k \psi \ln m} + 2\psi \ln m\right) \leq e^{-\psi \ln m} = m^{-\psi}.\label{eq:chi-square-bound}
\end{equation}
Using (\ref{eq:chi-square-bound}) in (\ref{eq:max-union-bound-overfit-N}) we get
\begin{equation}
    \Pr\left(Z_\text{max} \leq k +2\sqrt{k \psi \ln m} + 2\psi \ln m\right) \geq 1-\frac{1}{m^{\psi-1}}.
\end{equation}
Therefore, $ Z_\text{max} \leq  k + 2\sqrt{k\psi \ln m} + 2\psi \ln m$ with probability approaching one as $m\to \infty$ if $\psi>1$.

\begin{lemma}\label{lemma-Gaussian-bound}
Let $X_\text{max} = \underset{i}{\max}\big \{X_i\big \}_{i=1}^{m}$ where $X_1,X_2,\ldots,X_m$ is a sequence of identically distributed random variables (not necessarily independent) having a Gaussian distribution with zero mean and variance one. Then $X_\text{max} \leq \sqrt{2\ln m}$ with probability approaching one as $m \to \infty$.
\end{lemma}

\textit{Proof:} From the union bound we have
\begin{equation}
    \Pr \left(X_\text{max} \leq \eta\right) \geq 1-m\Pr \left(X_i \geq \eta\right).\label{eq:max-union-bound-Gaussian}
\end{equation}
Since $X_i \sim \mathcal{N}(0,1)$, from the Gaussian tail bound we have
\begin{equation}
    \Pr \left(X_i\geq \eta \right) \leq \frac{1}{\eta}\frac{e^{-\eta^2/2}}{\sqrt{2\pi}},\label{eq:Gaussian-tail-bound}
\end{equation}
for all $\eta >0$. Setting $\eta=\sqrt{2\ln m}$ in (\ref{eq:Gaussian-tail-bound}) we get
\begin{equation}
    \Pr \left(X_i\geq \sqrt{2\ln m} \right) \leq \frac{m^{-1}}{2\sqrt{\pi\ln m}} . \label{eq:Gaussian-tail-bound-misfit-N}
\end{equation}
Using (\ref{eq:Gaussian-tail-bound-misfit-N}) in (\ref{eq:max-union-bound-Gaussian}) we get
\begin{equation}
    \Pr \left(X_\text{max} \leq \sqrt{2\ln m} \right) \geq 1-\frac{1}{2\sqrt{\pi\ln m}}.
\end{equation}
Therefore, $X_\text{max} \leq \sqrt{2\ln m}$ with probability approaching one as $m \to \infty$.

\begin{lemma}\label{lemma-identifiability}
For any arbitrary support $\mathcal{I}\in \mathcal{I}^k_m \in \mathbb{M}$, under the asymptotic identifiability condition in (\ref{eq:identifiability-condition-on-A}) the following inequality holds
\begin{equation*}
    \left\lVert \mathbf{\Pi}^\perp_\mathcal{I}\mathbf{A}_\mathcal{S}\mathbf{x}_\mathcal{S} \right\rVert^2_2 >0.
\end{equation*}
\end{lemma}

\textit{Proof:} Let $\mathcal{S}' = \{ \mathcal{S}\setminus \mathcal{I} \}$. The true support $\mathcal{S}$ can be split into two disjoint subsets as $\mathcal{S} = \{\mathcal{S}\cap \mathcal{I}\} \cup \{ \mathcal{S}\setminus\mathcal{I}\}$. Since $\text{span}(\mathbf{A}_{\mathcal{S}\cap \mathcal{I}}) \subset \text{span}(\mathbf{A}_\mathcal{I})$ we have
\begin{align*}
        \lVert \mathbf{\Pi}^\perp_\mathcal{I}\mathbf{A}_\mathcal{S}\mathbf{x}_\mathcal{S} \rVert^2_2 = & \lVert \mathbf{\Pi}^\perp_\mathcal{I}\mathbf{A}_{\mathcal{S}'}\mathbf{x}_{\mathcal{S}'}\rVert^2_2 \\
        = & N\mathbf{x}_{\mathcal{S}'}^T\left(N^{-1}\mathbf{A}_{\mathcal{S}'}^T\mathbf{\Pi}^\perp_\mathcal{I}\mathbf{A}_{\mathcal{S}'}\right)\mathbf{x}_{\mathcal{S}'}.
\end{align*}
Now, consider the matrix $\mathbf{M} = \begin{bmatrix}
    \mathbf{A}_{\mathcal{S}'} & \mathbf{A}_{\mathcal{I}}
    \end{bmatrix}$
where $card(\mathcal{S}')\leq K$ and $card(\mathcal{I})\leq K$, such that $card(\mathcal{S}'\cup \mathcal{I})\leq 2K$. Under the assumption (\ref{eq:identifiability-condition-on-A}) 
\begin{align}
    N^{-1}\mathbf{M}^T\mathbf{M} 
    =N^{-1}\begin{bmatrix}
    \mathbf{A}^T_{\mathcal{S}'}\mathbf{A}_{\mathcal{S}'} & \mathbf{A}^T_{\mathcal{S}'} \mathbf{A}_{\mathcal{I}}\\
    \mathbf{A}^T_{\mathcal{I}}\mathbf{A}_{\mathcal{S}'} & \mathbf{A}^T_{\mathcal{I}}\mathbf{A}_{\mathcal{I}}
    \end{bmatrix}
\end{align}
is a bounded positive definite matrix.
Then the Schur complement of the block matrix $\mathbf{A}_\mathcal{I}^T\mathbf{A}_\mathcal{I}$ is
\begin{align*}
    &N^{-1}\left[\mathbf{A}^T_{\mathcal{S}'}\mathbf{A}_{\mathcal{S}'}-\mathbf{A}^T_{\mathcal{S}'}\mathbf{A}_{\mathcal{I}}(\mathbf{A}^T_{\mathcal{I}}\mathbf{A}_{\mathcal{I}})^{-1}\mathbf{A}^T_{\mathcal{I}}\mathbf{A}_{\mathcal{S}'}\right]\\
    = &N^{-1}\mathbf{A}^T_{\mathcal{S}'}\mathbf{\Pi}^\perp_\mathcal{I} \mathbf{A}_{\mathcal{S}'}
\end{align*}
is also positive definite and bounded as $N \to \infty$.
Let $\widetilde{\mathbf{M}} = N^{-1}\mathbf{A}_{\mathcal{S}'}^T\mathbf{\Pi}^\perp_\mathcal{I}\mathbf{A}_{\mathcal{S}'}$, then, $ \mathbf{x}_{\mathcal{S}'}^T \widetilde{\mathbf{M}}\mathbf{x}_{\mathcal{S}'} =b\text{ (say) }=\mathcal{O}(1)>0$. Hence, $\lVert\mathbf{\Pi}^\perp_\mathcal{I}\mathbf{A}_\mathcal{S}\mathbf{x}_\mathcal{S} \rVert^2_2 = Nb >0$ for all $\mathcal{I} \in \mathcal{I}^k_m \in \mathbb{M}$.



\bibliographystyle{IEEEtran}
\bibliography{main.bib}

\begin{thebibliography}{10}
\providecommand{\url}[1]{#1}
\csname url@samestyle\endcsname
\providecommand{\newblock}{\relax}
\providecommand{\bibinfo}[2]{#2}
\providecommand{\BIBentrySTDinterwordspacing}{\spaceskip=0pt\relax}
\providecommand{\BIBentryALTinterwordstretchfactor}{4}
\providecommand{\BIBentryALTinterwordspacing}{\spaceskip=\fontdimen2\font plus
\BIBentryALTinterwordstretchfactor\fontdimen3\font minus
  \fontdimen4\font\relax}
\providecommand{\BIBforeignlanguage}[2]{{%
\expandafter\ifx\csname l@#1\endcsname\relax
\typeout{** WARNING: IEEEtran.bst: No hyphenation pattern has been}%
\typeout{** loaded for the language `#1'. Using the pattern for}%
\typeout{** the default language instead.}%
\else
\language=\csname l@#1\endcsname
\fi
#2}}
\providecommand{\BIBdecl}{\relax}
\BIBdecl

\bibitem{MOS_overview_2018}
J.~Ding, V.~Tarokh, and Y.~Yang, ``Model selection techniques: An overview,''
  \emph{IEEE Signal Processing Magazine}, vol.~35, no.~6, pp. 16--34, 2018.

\bibitem{MOS_review_Stoica2004}
P.~Stoica and Y.~Selen, ``Model-order selection: a review of information
  criterion rules,'' \emph{IEEE Signal Processing Magazine}, vol.~21, no.~4,
  pp. 36--47, 2004.

\bibitem{MS_Rao_2001}
C.~Rao, Y.~Wu, S.~Konishi, and R.~Mukerjee, ``On model selection,''
  \emph{Lecture Notes-Monograph Series}, pp. 1--64, 2001.

\bibitem{MS_review_Anderson_2004}
D.~Anderson and K.~Burnham, ``Model selection and multi-model inference,''
  \emph{Second. NY: Springer-Verlag}, vol.~63, p.~10, 2004.

\bibitem{MS_review_chakrbarti_2011}
A.~Chakrabarti and J.~K. Ghosh, ``{AIC}, {BIC} and recent advances in model
  selection,'' \emph{Philosophy of statistics}, pp. 583--605, 2011.

\bibitem{AIC_Akaike1974}
H.~Akaike, ``A new look at the statistical model identification,'' \emph{IEEE
  transactions on automatic control}, vol.~19, no.~6, pp. 716--723, 1974.

\bibitem{BIC_Schwarz1978}
G.~Schwarz \emph{et~al.}, ``Estimating the dimension of a model,'' \emph{Annals
  of statistics}, vol.~6, no.~2, pp. 461--464, 1978.

\bibitem{MDL1978}
J.~Rissanen, ``Modeling by shortest data description,'' \emph{Automatica},
  vol.~14, no.~5, pp. 465--471, 1978.

\bibitem{gMDL_Hansen2001}
M.~H. Hansen and B.~Yu, ``Model selection and the principle of minimum
  description length,'' \emph{Journal of the American Statistical Association},
  vol.~96, no. 454, pp. 746--774, 2001.

\bibitem{NML_Rissanen2000}
J.~Rissanen, ``{MDL} denoising,'' \emph{IEEE Transactions on Information
  Theory}, vol.~46, no.~7, pp. 2537--2543, 2000.

\bibitem{PAL_Stoica2013}
P.~Stoica and P.~Babu, ``Model order estimation via penalizing adaptively the
  likelihood {(PAL)},'' \emph{Signal Processing}, vol.~93, no.~11, pp.
  2865--2871, 2013.

\bibitem{EBIC2008}
J.~Chen and Z.~Chen, ``Extended {Bayesian} information criteria for model
  selection with large model spaces,'' \emph{Biometrika}, vol.~95, no.~3, pp.
  759--771, 2008.

\bibitem{EFIC2018}
A.~Owrang and M.~Jansson, ``A model selection criterion for high-dimensional
  linear regression,'' \emph{IEEE Transactions on Signal Processing}, vol.~66,
  no.~13, pp. 3436--3446, 2018.

\bibitem{EEF_Kay2005}
S.~Kay, ``Exponentially embedded families-new approaches to model order
  estimation,'' \emph{IEEE Transactions on Aerospace and Electronic Systems},
  vol.~41, no.~1, pp. 333--345, 2005.

\bibitem{FIC}
H.~Bozdogan, ``Model selection and {Akaike's} information criterion {(AIC)}:
  The general theory and its analytical extensions,'' \emph{Psychometrika},
  vol.~52, no.~3, pp. 345--370, 1987.

\bibitem{CV_picard1984}
R.~R. Picard and R.~D. Cook, ``Cross-validation of regression models,''
  \emph{Journal of the American Statistical Association}, vol.~79, no. 387, pp.
  575--583, 1984.

\bibitem{CV_p_greater_N}
L.~de~Torrent{\'e} and T.~Hastie, ``Does cross-validation work when $p\gg n$?''
  2012.

\bibitem{RRT-2018}
S.~Kallummil and S.~Kalyani, ``Signal and noise statistics oblivious orthogonal
  matching pursuit,'' in \emph{International Conference on Machine
  Learning}.\hskip 1em plus 0.5em minus 0.4em\relax PMLR, 2018, pp. 2429--2438.

\bibitem{gohain2020_MBT}
P.~B. Gohain and M.~Jansson, ``Relative cost based model selection for sparse
  high-dimensional linear regression models,'' in \emph{ICASSP IEEE
  International Conference on Acoustics, Speech and Signal Processing
  (ICASSP)}.\hskip 1em plus 0.5em minus 0.4em\relax IEEE, 2020, pp. 5515--5519.

\bibitem{OMP_Cai2011}
T.~T. Cai and L.~Wang, ``Orthogonal matching pursuit for sparse signal recovery
  with noise,'' \emph{IEEE Transactions on Information theory}, vol.~57, no.~7,
  pp. 4680--4688, 2011.

\bibitem{Gohain2022EBICR}
P.~B. Gohain and M.~Jansson, ``New improved criterion for model selection in
  sparse high-dimensional linear regression models,'' in \emph{ICASSP IEEE
  International Conference on Acoustics, Speech and Signal Processing
  (ICASSP)}, 2022, pp. 5692--5696.

\bibitem{S_Kay_estimation_book}
S.~M. Kay, \emph{Fundamentals of statistical signal processing: estimation
  theory}.\hskip 1em plus 0.5em minus 0.4em\relax Prentice Hall PTR, 1993.

\bibitem{Stoica_BIC_SNR}
P.~Stoica and P.~Babu, ``On the proper forms of {BIC} for model order
  selection,'' \emph{IEEE Transactions on Signal Processing}, vol.~60, no.~9,
  pp. 4956--4961, 2012.

\bibitem{GOHAIN-BIC-R}
P.~B. Gohain and M.~Jansson, ``Scale-invariant and consistent {Bayesian}
  information criterion for order selection in linear regression models,''
  \emph{Signal Processing}, p. 108499, 2022.

\bibitem{Schmidt_2011_Consistency_of_MDL}
D.~F. Schmidt and E.~Makalic, ``The consistency of {MDL} for linear regression
  models with increasing signal-to-noise ratio,'' \emph{IEEE transactions on
  signal processing}, vol.~60, no.~3, pp. 1508--1510, 2011.

\bibitem{djuric1998asymptotic}
P.~M. Djuric, ``Asymptotic {MAP} criteria for model selection,'' \emph{IEEE
  Transactions on Signal Processing}, vol.~46, no.~10, pp. 2726--2735, 1998.

\bibitem{inconsistency_of_BIC_Kay_2011}
Q.~Ding and S.~Kay, ``Inconsistency of the {MDL}: On the performance of model
  order selection criteria with increasing signal-to-noise ratio,'' \emph{IEEE
  Transactions on Signal Processing}, vol.~59, no.~5, pp. 1959--1969, 2011.

\bibitem{zhang2008sparsity}
C.-H. Zhang and J.~Huang, ``The sparsity and bias of the lasso selection in
  high-dimensional linear regression,'' \emph{The Annals of Statistics},
  vol.~36, no.~4, pp. 1567--1594, 2008.

\bibitem{meinshausen2006high}
N.~Meinshausen and P.~B{\"u}hlmann, ``High-dimensional graphs and variable
  selection with the lasso,'' \emph{The annals of statistics}, vol.~34, no.~3,
  pp. 1436--1462, 2006.

\bibitem{lasso_tibshirani1996}
R.~Tibshirani, ``Regression shrinkage and selection via the lasso,''
  \emph{Journal of the Royal Statistical Society: Series B (Methodological)},
  vol.~58, no.~1, pp. 267--288, 1996.

\bibitem{LARS_Efron2004}
B.~Efron, T.~Hastie, I.~Johnstone, and R.~Tibshirani, ``Least angle
  regression,'' \emph{The Annals of statistics}, vol.~32, no.~2, pp. 407--499,
  2004.

\bibitem{mathai1992quadratic}
A.~M. Mathai and S.~B. Provost, \emph{Quadratic forms in random variables:
  theory and applications}.\hskip 1em plus 0.5em minus 0.4em\relax Dekker,
  1992.

\bibitem{laurent2000adaptive}
B.~Laurent and P.~Massart, ``Adaptive estimation of a quadratic functional by
  model selection,'' \emph{Annals of Statistics}, pp. 1302--1338, 2000.

\end{thebibliography}
\ifCLASSOPTIONcaptionsoff
  \newpage
\fi

\end{document}